\documentclass[a4paper,fleqn,usenatbib]{mnras}
\usepackage[T1]{fontenc}
\usepackage{ae,aecompl}
\usepackage{newtxtext,newtxmath}
\usepackage{graphicx}	
\usepackage{amsmath}	
\usepackage{amssymb}	

\title[ALMA maps of HCO towards IRAS 16293$-$2422]
{First ALMA maps of HCO, an important precursor of complex organic molecules, towards IRAS 16293$-$2422 }

\author[V. M. Rivilla et al.]
{V. M. Rivilla$^{1}$\thanks{E-mail: rivilla@arcetri.astro.it}, 
M. T. Beltr\'an$^{1}$,
A. Vasyunin$^{2,3,4}$, 
P. Caselli$^{2}$,
S. Viti$^{5}$,
\newauthor{F. Fontani$^{1}$,
and R. Cesaroni$^{1}$}
\\
$^{1}$INAF-Osservatorio Astrofisico di Arcetri, Largo Enrico Fermi 5, I-50125, Florence, Italy\\
$^{2}$Max-Planck-Institute for Extraterrestrial Physics, Garching, Germany\\ 
$^{3}$Ural Federal University, Ekaterinburg, Russia\\
$^{4}$Visiting Leading Researcher, Engineering Research Institute 'Ventspils International Radio
Astronomy Centre' of Ventspils University of Applied Sciences, \\
 In\v{z}enieru 101, Ventspils LV-3601, Latvia \\
$^{5}$Department of Physics and Astronomy, University College London, Gower street, London, WC1E 6BT, UK\\
}

\date{Accepted XXX. Received YYY; in original form ZZZ}

\pubyear{2017}

\begin{document}
\label{firstpage}
\pagerange{\pageref{firstpage}--\pageref{lastpage}}
\maketitle

\begin{abstract}
The formyl radical HCO has been proposed as the basic precursor of many complex organic molecules such as methanol (CH$_3$OH) or glycolaldehyde (CH$_2$OHCHO). Using ALMA, we have mapped, for the first time at high angular resolution ($\sim$1$^{\prime\prime}$, $\sim$140 au), HCO towards the Solar-type protostellar binary IRAS 16293$-$2422, where numerous complex organic molecules have been previously detected. We also detected several lines of the chemically related species H$_2$CO, CH$_3$OH and CH$_2$OHCHO. The observations revealed compact HCO emission arising from the two protostars. The line profiles also show redshifted absorption produced by foreground material of the circumbinary envelope that is infalling towards the protostars. Additionally, IRAM 30m single-dish data revealed a more extended HCO component arising from the common circumbinary envelope. The comparison between the observed molecular abundances and our chemical model suggests that whereas the extended HCO from the envelope can be formed via gas-phase reactions during the cold collapse of the natal core, the HCO in the hot corinos surrounding the protostars is predominantly formed by the hydrogenation of CO on the surface of dust grains and subsequent thermal desorption during the protostellar phase. 
The derived abundance of HCO in the dust grains is high enough to produce efficiently more complex species such as H$_2$CO, CH$_3$OH, and CH$_2$OHCHO by surface chemistry. We found that the main formation route of CH$_2$OHCHO is the reaction between HCO and CH$_2$OH.
\end{abstract}





\section{Introduction}
\label{intro}


The formation of complex organic molecules (COMs) $-$ carbon-based compounds with more than 5 atoms (\citealt{herbst2009}) $-$ is being intensively debated in astrochemistry. 
COMs play a central role in prebiotic chemistry and may be directly linked to the origin of life (e.g. \citealt{caselli2012}). 
Numerous efforts have been done in the last years to understand how COMs are formed in the insterstellar medium (ISM), by combining observations (e.g., \citealt{belloche2009,jorgensen2016,martin-domenech2017,codella2017,rivilla2017a,rivilla2017b}), chemical modeling (e.g., \citealt{garrod08,vasyunin2013a,vasyunin2013b,balucani15,taquet2016,vasyunin2017,coutens2018,bergantini2018,quenard2018a}), and laboratory experiments (e.g. \citealt{fedoseev15,chuang2016,chuang2017}).
However, despite all efforts, our understanding about the synthesis of COMs in the ISM is still very limited.
Two general paradigms have been proposed: i) gas-phase chemistry triggered by the evaporation (thermal or non-thermal) of interstellar ices (e.g., \citealt{millar91}; \citealt{vasyunin2013b}; \citealt{balucani15}; \citealt{vasyunin2017}); ii) hydrogenation and/or radical-radical reactions on dust grain surfaces (e.g., \citealt{garrod2006,garrod08}). 

A key step to understand how complex molecules are built up in the ISM is to study their molecular precursors, that is, the basic pieces that lead to their formation. Many chemical models and laboratory experiments have proposed that the simple formyl radical, HCO, is the precursor of COMs like e.g. methanol (\citealt{watanabe2002}), the sugar-like molecule glycolaldehyde and the sugar-alcohol ethylene glycol (\citealt{bennett07}; \citealt{woods12,woods13}; \citealt{fedoseev15}; \citealt{butscher15}; \citealt{chuang2017}),  formamide (\citealt{jones2011}, \citealt{fedoseev2016}), and N-methylformamide (\citealt{belloche2017}). 
However, despite the importance of HCO to build-up chemical complexity, little is known so far about its formation itself. Two main scenarios have been proposed: i) formation on the surface of dust grains via neutral$-$neutral reactions at cold starless stages (e.g., \citealt{tielens1982}; \citealt{brown1988}; \citealt{dartois1999}; \citealt{watanabe2002}; \citealt{woon2002}; \citealt{garrod08}; \citealt{bacmann2016}); and ii) formation via gas-phase chemistry (\citealt{bacmann2016}; \citealt{hickson2016}).

\begin{table}
\tabcolsep 5.0pt
\begin{center}
\caption{ALMA observations used in this work.}
\label{table-observations}
\begin{tabular}{c c c c c }
\hline
spw & Frequency range      & Synthesized beam	    & $\Delta {\rm v}$            &  rms	       \\ 
 & (GHz)           &   ($\arcsec\times\arcsec$)     & (km s$^{-1}$) & (mJy)          \\ 
\hline
0     &  86.58-87.05    & 1.45$\times$1.04  & 0.84  &  2.0 \\
1     &  88.50-88.97    & 1.41$\times$1.02  & 0.83  &  2.0 \\
2     &  99.00-99.23    & 1.28$\times$0.92  & 0.74  &  1.4 \\
3     &  101.10-101.57  & 1.25$\times$0.90  & 0.72  &  2.3 \\
\hline
\end{tabular}
\end{center}
\end{table}









To understand how HCO is formed, and what is its role in the formation of more complex species, dedicated observations of this molecule are needed. HCO has been detected in very different environments: diffuse clouds (\citealt{liszt2014}), molecular clouds (\citealt{snyder1976,snyder1985}; \citealt{schenewerk1986,schenewerk1988}), photon-dominated regions (\citealt{schilke2001}; \citealt{gerin2009}), cold dark clouds (\citealt{cernicharo2012}; \citealt{agundez2015}), shocks (\citealt{jimenez-serra2004}), starless cores (\citealt{frau2012}; \citealt{bacmann2016}; \citealt{spezzano2017}),
low-mass protostellar objects (\citealt{caux2011}; \citealt{bacmann2016}) and massive star-forming regions (\citealt{sanchez-monge2013}; Rivilla et al., in prep.). 
However, these observations have been carried out with single-dish telescopes, and hence, up to now, no high angular resolution observations of HCO are available.
This has prevented us from revealing the HCO spatial distribution and properly deriving source-average abundances to be compared with chemical models.

\begin{table*}
\tabcolsep 5.0pt
\begin{center}
\caption{Molecular transitions of the different species that are clearly unblended towards IRAS16293 source B.}
\label{table-transitions}
\begin{tabular}{l c c c c}
\hline
Molecule & Frequency       & Transition	                &  log A$_{ul}$ 	& E$_{\rm up}$   	           \\ 
& (GHz)           &         & (s$^{-1}$) & (K)             \\ 
\hline
HCO     &  86.67076     &  1$_{0,1}-$0$_{0,0}$, J=3/2--1/2, F=2--1  &  -5.3289   & 4   \\
HCO     &  86.70836     &  1$_{0,1}-$0$_{0,0}$, J=3/2--1/2, F=1--0  &  -5.3377   &  4  \\
HCO     &  86.77746     &  1$_{0,1}-$0$_{0,0}$, J=1/2--1/2, F=1--1     &  -5.3366    & 4   \\
HCO     &  86.80578     &  1$_{0,1}-$0$_{0,0}$, J=1/2--1/2, F=0--1     &  -5.3268    &  4   \\
\hline
H$_2$CO & 101.33299     &  6$_{1,5}-$6$_{1,6}$     &  -5.8038    &  88    \\
\hline
CH$_3$OH & 86.61560     &  7$_{2,6}-$6$_{3,3}$ - -     &  -6.1646    &   103  \\
CH$_3$OH & 86.90295     &  7$_{2,5}$- 6$_{3,4}$ + +     &  -6.1596    &  103   \\
CH$_3$OH & 88.59479     &  15$_{3,13}$ - 14$_{4,10}$ ++    & -5.9593     &  328   \\
CH$_3$OH & 88.94009     &  15$_{3,12}$ - 14$_{4,11}$ - -     &  -5.9539    &  328   \\
\hline
CH$_2$OHCHO & 86.60057      &   17$_{5,2}$ - 17$_{4,13}$     &   -4.4770   &  101  \\
CH$_2$OHCHO$^{(a)}$ &  86.86239     &  7$_{4,3}$ - 7$_{3,4}$    &  -4.9138     &  25  \\
CH$_2$OHCHO &  86.87650     &    $20_{4,16}$ - 20$_{3,17}$      &   -4.8626  &  427  \\
CH$_2$OHCHO$^{(a)}$ & 88.53041      &   8$_{4,5}$ - 8$_{3,6}$    & -4.8217     &  30   \\
CH$_2$OHCHO & 88.69126    &   12$_{3,10}$ - 12$_{2,11}$     & -4.7155       & 49   \\
CH$_2$OHCHO & 88.89245    &   9$_{4,6}$ - 9$_{3,7}$     &   -4.7558     &  35   \\
CH$_2$OHCHO &  99.06847   &   14$_{4,11}$ - 14$_{3,12}$     &  -4.5044    & 67  \\ 
CH$_2$OHCHO & 101.11631   &   21$_{4,17}$ - 21$_{3,18}$     &   -4.3682  &  143   \\
CH$_2$OHCHO$^{(a)}$ &  101.21981     &   $18_{3,15}$ - $18_{2,16}$     &  -4.5990    & 497   \\
CH$_2$OHCHO & 101.23217   &   15$_{2,13}$ - 15$_{1,14}$     &   -4.5990    & 71   \\
CH$_2$OHCHO$^{(a)}$ &  101.51469     &   $14_{3,12}$ - $14_{2,13}$     &  -5.2467    &  535 \\
CH$_2$OHCHO & 101.52785   &    14$_{5,9}$ - 14$_{4,10}$    &  -4.73347    &  74  \\
\hline
\end{tabular}
\end{center}
(a) Transitions used to perform AUTOFIT in source B.
\end{table*}







IRAS 16293$-$2422 (hereafter IRAS16293) is a low-mass protostellar system located at a distance of 141$^{+31}_{-21}$ pc (\citealt{dzib2018}).
It is composed of two Solar-like protostars, sources A and B, surrounded by chemically rich hot corinos (\citealt{cazaux2003,bottinelli2004}) separated in the plane of the sky by $\sim$5$\arcsec$ ($\sim$705 au), and whose masses are $\sim$0.5 $M_{\odot}$ (\citealt{looney2000}). Many COMs have been detected towards the two hot corinos of this system, including glycolaldehyde (CH$_2$OHCHO, \citealt{jorgensen12}), ethylene glycol ((CH$_2$OH)$_{2}$, \citealt{jorgensen12}), formamide (NH$_2$CHO, \citealt{kahane2013,coutens2016}) and methyl isocyanate (CH$_3$NCO, \citealt{martin-domenech2017}; \citealt{ligterink2017}). Their emission exhibits line profiles with linewidths of up to 8 km$^{-1}$ for A and $<$ 2 km s$^{-1}$ for B, due to the different inclination of the sources (B almost face-on). Due to its chemical richness, IRAS16293 is an excellent laboratory for astrochemical studies.

In this paper, we present for the first time interferometric ALMA observations of HCO towards IRAS16293. We study the spatial distribution of this molecule and compare its molecular abundance with that of more complex species: H$_2$CO, CH$_3$OH, and CH$_2$OHCHO. We also present different chemical models to understand the formation of HCO itself, and its role in the formation of more complex species such as glycolaldehyde.

\section{Observations}
\label{observations}

We carried out interferometric observations using 40 antennas of the Atacama Large Millimeter/Submillimeter Array (ALMA) in Cycle 3 in June 11 2016 as part of the project 2015.1.01193.S (PI Rivilla). The observations were performed in Band 3 (3 mm) with the array in a configuration with baselines ranging from 15\,m to 783\,m. The digital correlator was configured in four different spectral windows (spw) to cover lines of HCO, H$_2$CO, CH$_3$OH and CH$_2$OHCHO. The precipitable water vapor (pwv) during the observations was in the range 2.0$-$2.4 mm. Flux calibration was obtained through observations of Titan. The phase was calibrated from interleaved observations of the quasar J1625$-$2527. The bandpass correction was obtained by observing the  BL Lacertae object J1517$-$2422. The on-source observing time was about 70 min.
The phase center was RA=16h 32m 22.62s,  DEC=-24$^{\circ}$ 28$^{\prime}$ 32.46$^{\prime\prime}$.


The data were calibrated and imaged using standard ALMA calibration scripts of the Common Astronomy Software Applications package (CASA)\footnote{https://casa.nrao.edu}. 
We imaged with CASA the datacubes including the continuum, and used them to perform further analysis (see Section \ref{results}). 
The frequency range, synthesized beams, velocity resolution and rms of the datacubes of the different spw's are summarized in Table \ref{table-observations}. 
The flux density calibration uncertainty is of 5$\%$, consistent with that of the ALMA band 7 observations of the same source carried out by \citet{jorgensen2016} and with that estimated from recent analysis of calibrators in bands 3 and 6 (\citealt{bonato2018} and references therein).
The analysis of the data was done with
MADCUBA\footnote{Madrid Data Cube Analysis on ImageJ is a software developed in the Center of Astrobiology (Madrid, INTA-CSIC) to visualize and analyze astronomical single spectra and datacubes (Mart\'in et al., in prep., \citealt{rivilla2016}). MADCUBA is available at http://cab.inta-csic.es/madcuba/MADCUBA\_IMAGEJ/ImageJMadcuba.html} software package.

We have complemented our interferometric ALMA data with publicly available single-dish IRAM 30m data from the TIMASSS survey consisting on a single-pointing towards IRAS16293  (see details in \citealt{caux2011}).





\begin{figure*}
\includegraphics[width=13.85cm]{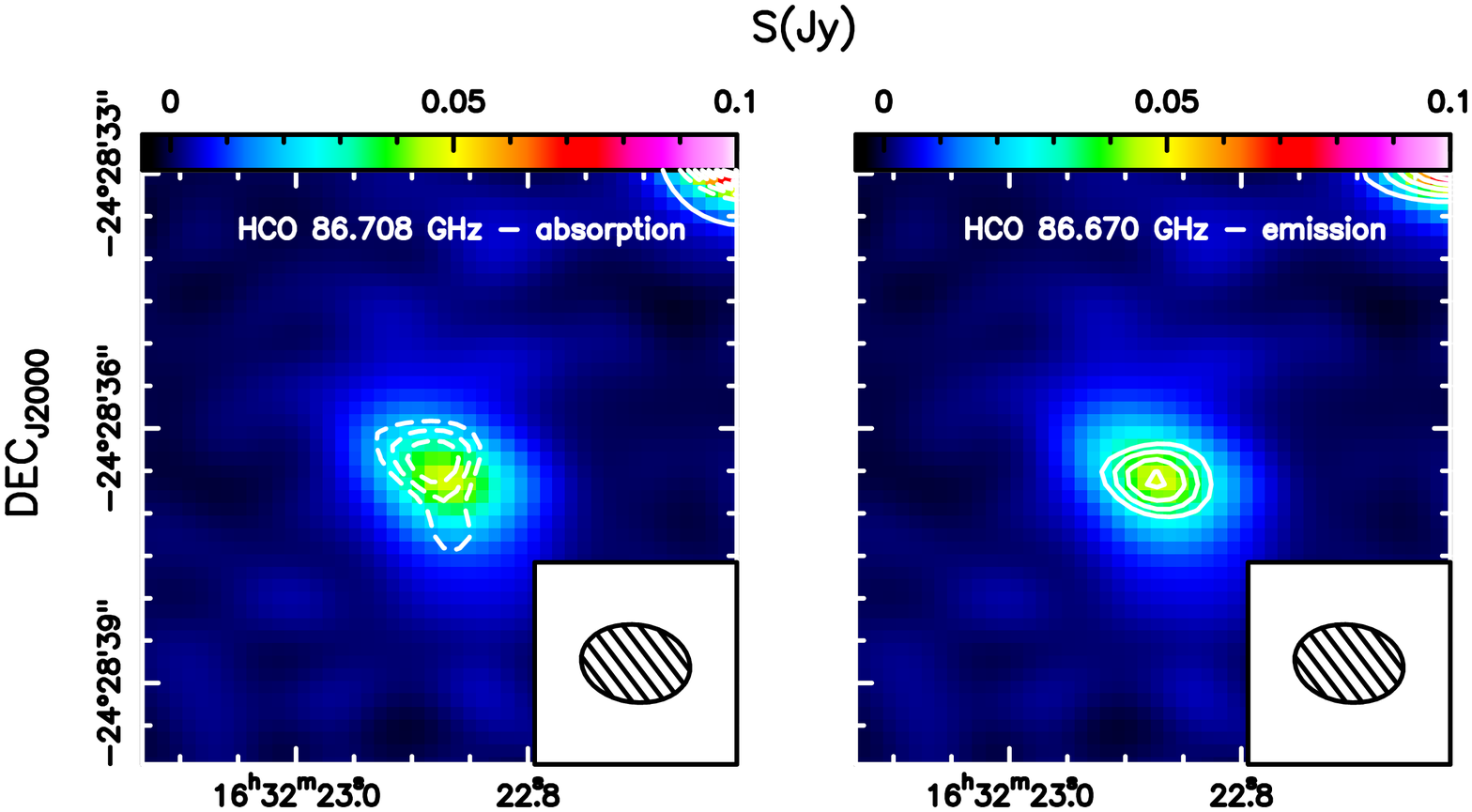}
\vskip3mm
\hspace{0mm}
\includegraphics[width=13.85cm]{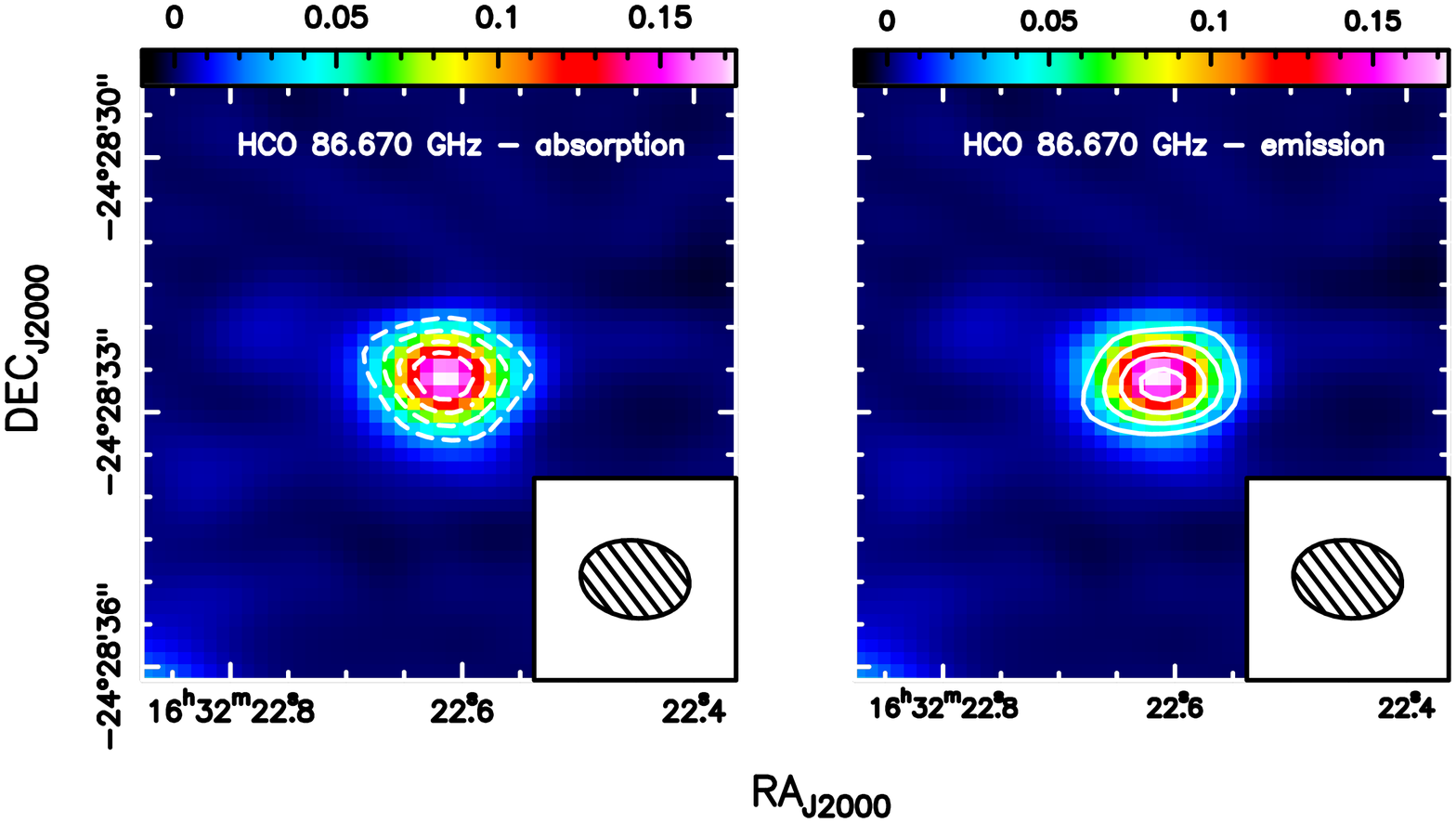}
 \caption{{\em Upper:} Integrated maps of HCO absorption at 86.708 GHz (velocity range between 3.5 and 4.5 km s$^{-1}$; left panel) and HCO emission at 86.670 GHz (velocity range between -0.5 and 4.3 km s$^{-1}$; right panel) towards IRAS16293 A. The absorption/emission contours start at -4.5/4.5 mJy, with steps of -1.5/1.5 mJy. The color scale is the 3 mm continuum emission in Jy. The beam is indicated in the lower right corner of each panel. {\em Lower}: Integrated maps of HCO absorption at 86.670 GHz (velocity range between 3.5 and 4.5 km s$^{-1}$; left panel) and HCO emission at 86.670 GHz (velocity range between 1.2 and 3.6 km s$^{-1}$; right panel) towards IRAS16293 B.  The absorption/emission contours start at -9/6 mJy, with steps of -6/3 mJy.}
    \label{fig-maps}
\end{figure*}

\begin{figure*}
\includegraphics[width=17cm]{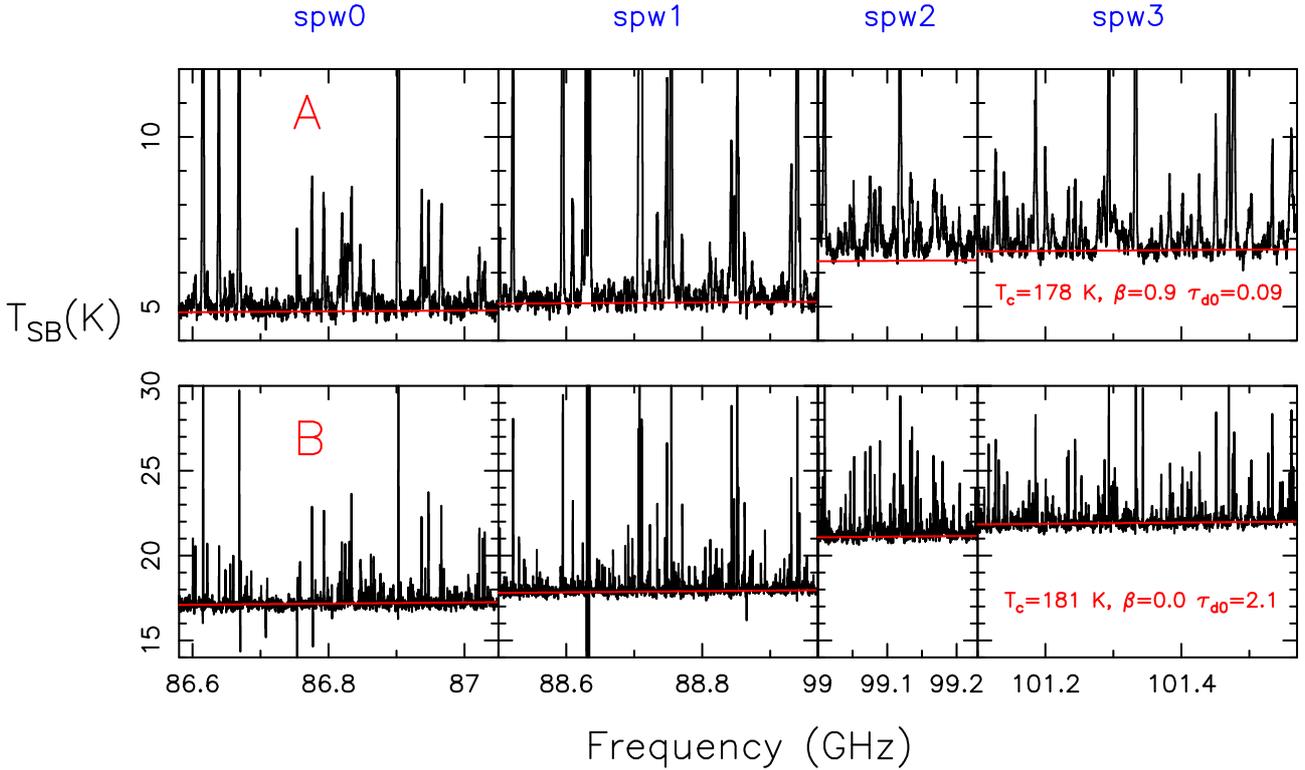}
 \caption{Spectra towards the continuum peak of IRAS16293 source A (upper) and IRAS16293 source B (lower) of the full bandwidth covered by the 4 different spectral windows. The red lines correspond to the fit of the continuum obtained using a modified black body function (see text). The parameters used in the fit are indicated in the right panels.}
    \label{fig-spws}
\end{figure*}

\begin{figure*}
\includegraphics[width=18cm]{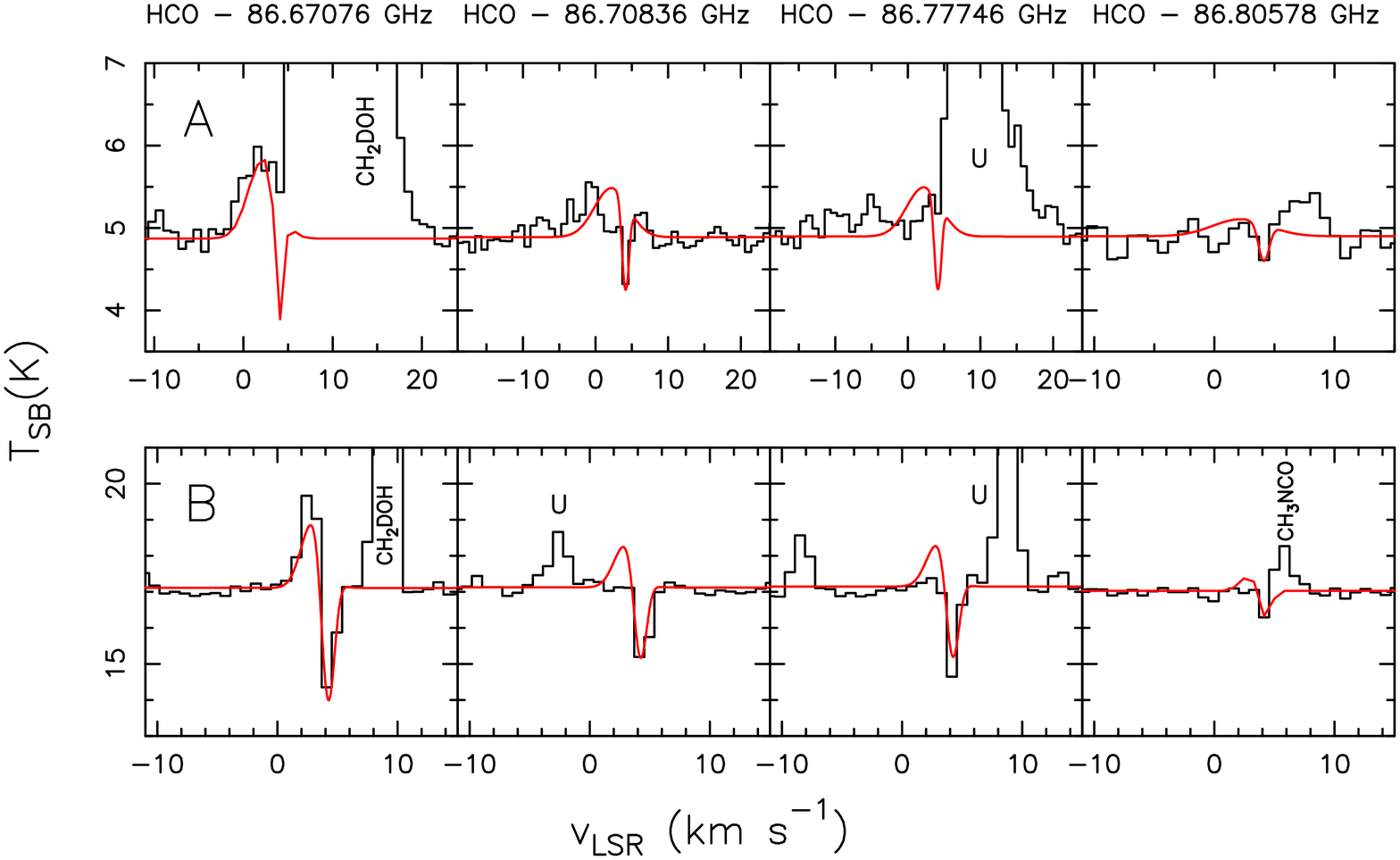}
 \caption{HCO spectra towards the continuum peak of IRAS16293 A (upper) and B (lower). The LTE best fits, including the continuum emission, are shown in red curves. Other molecular species in the spectra are also labeled with the their name or with an "U" if they remain unidentified.}
    \label{fig-HCO}
\end{figure*}

\section{Analysis}
\label{results}

\subsection{Continuum}
\label{analysis-continuum}

We show in Figure \ref{fig-maps} the 3mm continuum map towards the two hot corinos: sources A and B.
Figure \ref{fig-spws} shows the ALMA spectra towards the continuum peaks of the two sources. 
In Appendix \ref{continuum-fit} we explain in detail how we fitted the continuum emission of both hot corinos in all spectral windows by varying the values of  the dust emissivity spectral index $\beta$ and the dust optical depth $\tau_{d0}$.
The best solutions for the continuum fitting are $\beta$=0.9$\pm$0.4 and $\tau_{d0}$=0.09$\substack{+0.01 \\ -0.01}$ for source A, and $\beta$=0.0$\pm$0.5 and  $\tau_{d0}$=2.1$\substack{+0.6 \\ -0.4}$  for source B. The different values of $\beta$ might be due to several factors such as dust chemical composition, size distribution, porosity, geometry, or optical depth effects (see e.g. \citealt{ricci2012},\citealt{testi2014}).

The resulting continuum levels are shown with red lines in Figure \ref{fig-spws}. To create the continuum image we used this continuum level as a reference for selecting the line-free channels, which were used to substract the continuum in the uv-plane. The obtained continuum map at $\sim$94 GHz (Figure \ref{fig-maps}) has a synthesized beam of 1.09$\arcsec\times$0.78$\arcsec$. We obtained the deconvolved size of the continuum sources of A and B at 94 GHz by performing a two-dimensional Gaussian fitting: (1.1$\pm$0.1)$\arcsec\times$(0.7$\pm$0.1)$\arcsec$ (PA=45$\pm$9) and (0.46$\pm$0.03)$\arcsec\times$(0.42$\pm$0.04)$\arcsec$ (PA=148$\pm$42), respectively. These sizes are smaller than the synthesized beam of the continuum map, which means that the emission is not resolved. We used these sizes and the synthesized beams of each datacube (Table \ref{table-observations}) to calculate the filling factor  $f_{\rm c}(\nu)$ for the continuum fit (see Appendix \ref{continuum-fit}).

We used the derived optical depths $\tau_{d0}$ to calculate the molecular hydrogen column density of both hot corinos, using the expression $N_{\rm H_2}$ = $\tau_{d0}$ /($\mu  m_{\rm H} \kappa_{0}$), where $\mu$ is the mean molecular mass per hydrogen atom (2.8), m$_{\rm H}$ is the hydrogen mass and $ \kappa_{\nu}$ is the absorption coefficient per unit density at frequency $\nu_{0}$. In the calculation we assumed a gas-to-dust mass ratio of 100, and we calculated $ \kappa_{0}$  for each source extrapolating from the value at 1 mm of 0.009 cm$^2$ g$^{-1}$ (thin ices in a H$_2$ density of 10$^6$ cm$^{-3}$; see \citealt{ossenkopf94}) and using the derived values of $\beta$, which gives  0.0044 and 0.009 cm$^2$ g$^{-1}$ for source A and B, respectively. We obtained N(H$_2$)=4.4$\times$10$^{24}$ cm$^{-2}$ and N(H$_2$)=5.0$\times$10$^{25}$ cm$^{-2}$ for A and B, respectively.
These hydrogen column densities are in good agreement with previous estimates: 3.5$\times$10$^{24}$ cm$^{-2}$ for source A  (\citealt{bottinelli2004}), and ($>$1.2$-$2.8)$\times$10$^{25}$ cm$^{-2}$ (\citealt{bottinelli2004,jorgensen2016,martin-domenech2017}) for source B.

\subsection{Molecular lines}

\subsubsection{Identification}
\label{analysis-lines-identification}

The identification of the lines was performed using the SLIM (Spectral Line Identification and Modeling) tool of MADCUBA, which uses information from publicly available spectral catalogs. For the analysis of this work we used molecular entries from JPL\footnote{https://spec.jpl.nasa.gov} (\citealt{pickett1998}) and CDMS\footnote{https://www.astro.uni-koeln.de/cdms} (\citealt{muller2001,muller2005,endres2016}). We searched for the 3 mm quadruplet of HCO, and also for H$_2$CO, CH$_3$OH, and CH$_2$OHCHO lines. In Table \ref{table-transitions} we include the transitions of each molecule that are clearly unblended towards source B. 
In Appendix \ref{appendix-line-parameters} we present the line parameters (areas, intensities, v$_{\rm LSR}$ and FWHW) of the different molecular transitions studied in this work obtained from individual Gaussian fits.


HCO was detected, in emission and redshifted absorption, towards the positions of the two hot corinos. Figure \ref{fig-maps} shows the spatial distribution of HCO, overplotted on the continuum image. The maps show that HCO (the two components, emission and absorption) is not extended but compact in the two protostellar objects, and coincident with the continuum. The deconvolved sizes of the HCO emission and absorption derived by a two-dimensional Gaussian fitting are smaller than the synthesized beam of the datacubes. The same occurs also for H$_2$CO, CH$_3$OH and CH$_2$OHCHO. Therefore, their spatial distributions are not resolved. For simplicity, we will assume hereafter that the molecular emission/absorption share the same size of the continuum, obtained as $(\theta_{maj} {\rm x} \theta_{min})^{0.5}$. This gives 0.88$\arcsec$ for source A and 0.44$\arcsec$ for source B. We note that these sizes are very similar to those used in other works, e.g. 0.9$\arcsec$ for source A (\citealt{oya2016}) and 0.5$\arcsec$ for source B (\citealt{jorgensen2016,ligterink2017,martin-domenech2017}).



Figure \ref{fig-HCO} shows the spectra of the HCO quadruplet towards the hot corinos.
The brighter HCO line at 86.670 GHz presents a clear inverse P-Cygni profile towards source B, with emission at 2.5 km s$^{-1}$ and absorption at 4.3 km s$^{-1}$. 
The absorption is also evident in the other three HCO lines of the quadruplet (upper panels of Figure \ref{fig-HCO}). 
A similar P-Cygni profile was also observed in other molecular species, methyl formate (CH$_3$OCHO), by \citet{pineda2012}. 
In the case of source A, the HCO line at 86.670 GHz was also detected (see lower left panel in Figure \ref{fig-HCO}), although it is partially blended with a transition of CH$_2$DOH, due to the larger linewidths of this source. The absorption component is not clearly seen in this HCO transition, due to the contamination by CH$_2$DOH. However, the absorption is clear in the HCO transition at 86.70836 GHz (Figure \ref{fig-HCO}).

\begin{figure*}
\hspace{-1.2cm}
\includegraphics[width=16cm]{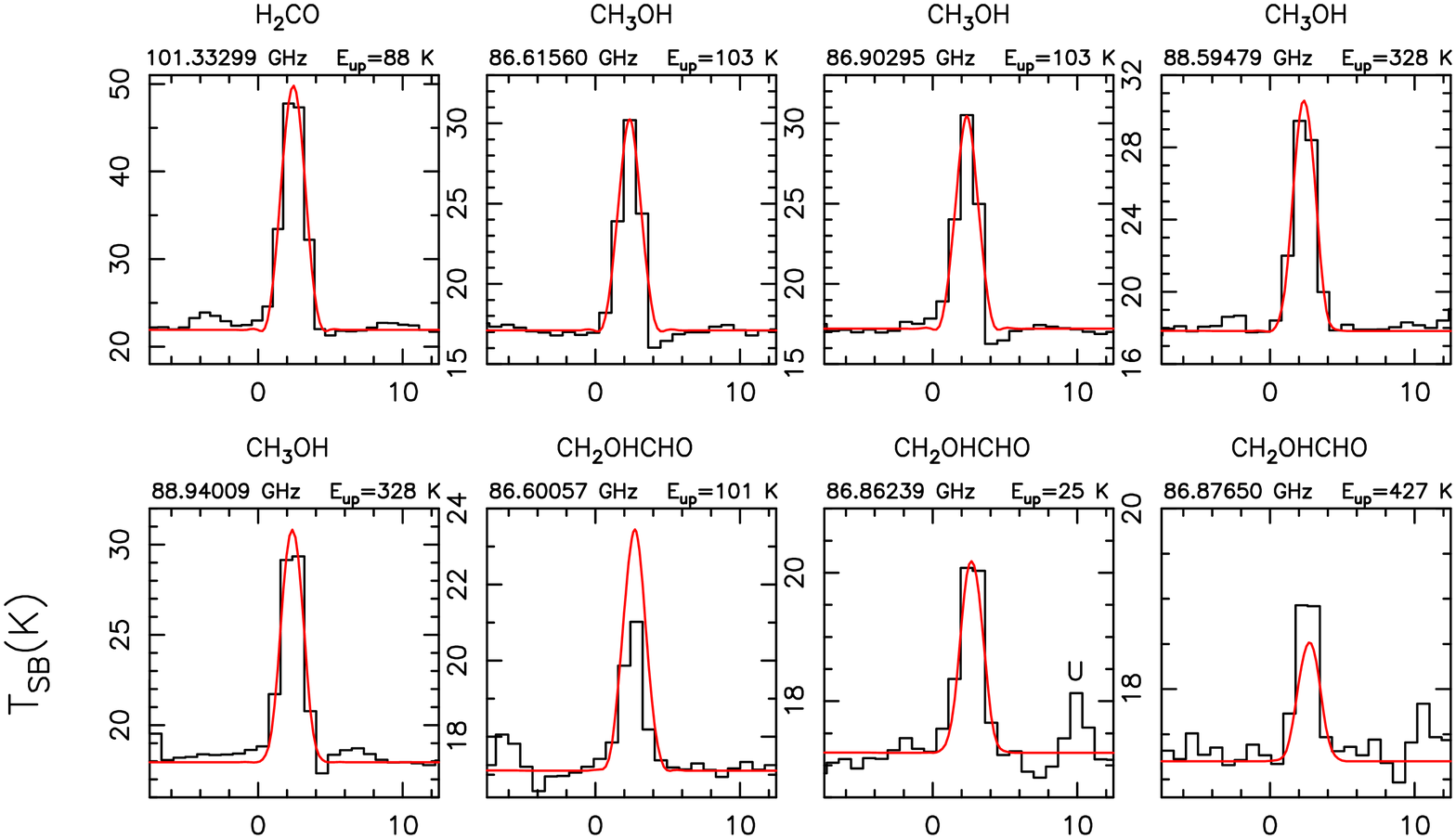}
\vskip4mm
\includegraphics[width=14.9cm]{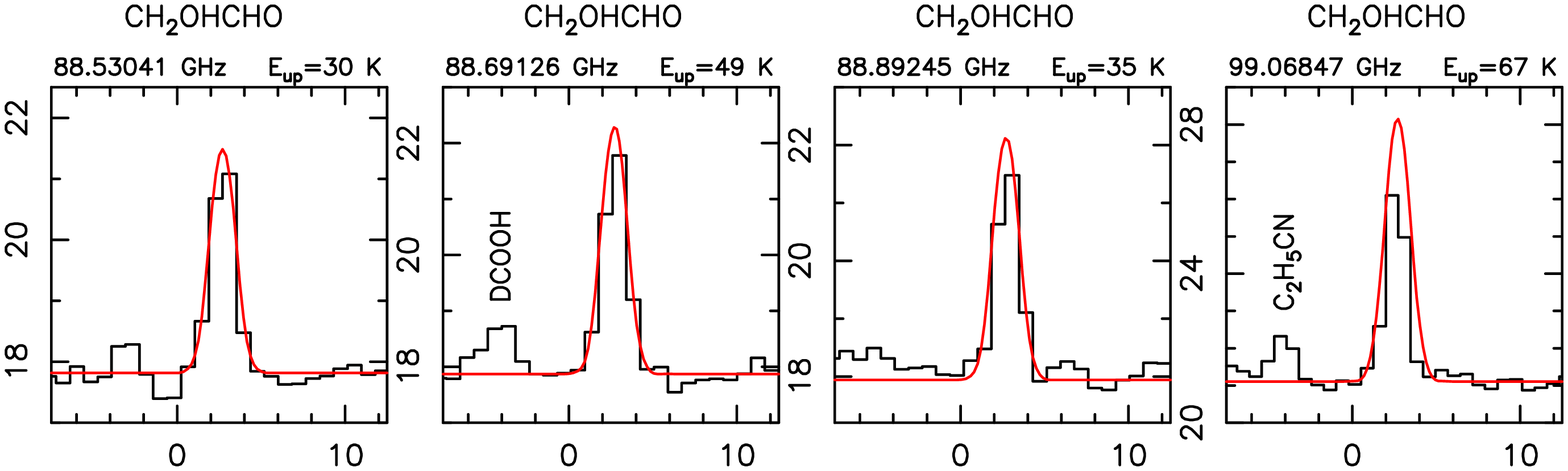}
\vskip4mm
\includegraphics[width=14.9cm]{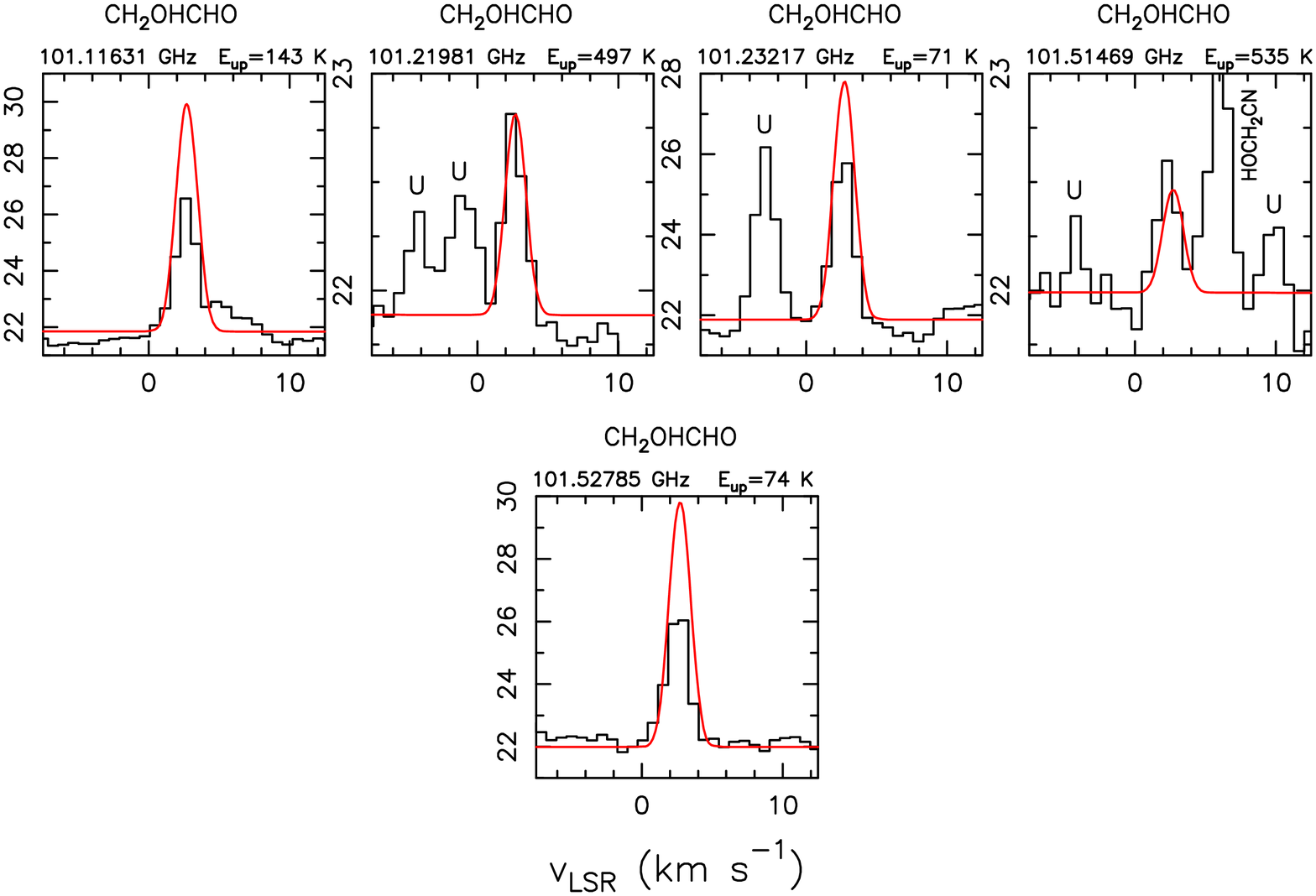}
 \caption{Unblended molecular transitions detected with ALMA of H$_2$CO, CH$_3$OH and CH$_2$OHCHO towards IRAS16293 source B. The MADCUBA-AUTOFIT LTE best fits, including the continuum, are shown in red. The frequency and $E_{\rm up}$ of each transition are indicated above each panel. Other molecular species in the spectra are also labeled with the their name or with an "U" if they remain unidentified.
 }
    \label{fig-molecules}
\end{figure*}
\begin{figure*}
\includegraphics[width=16cm]{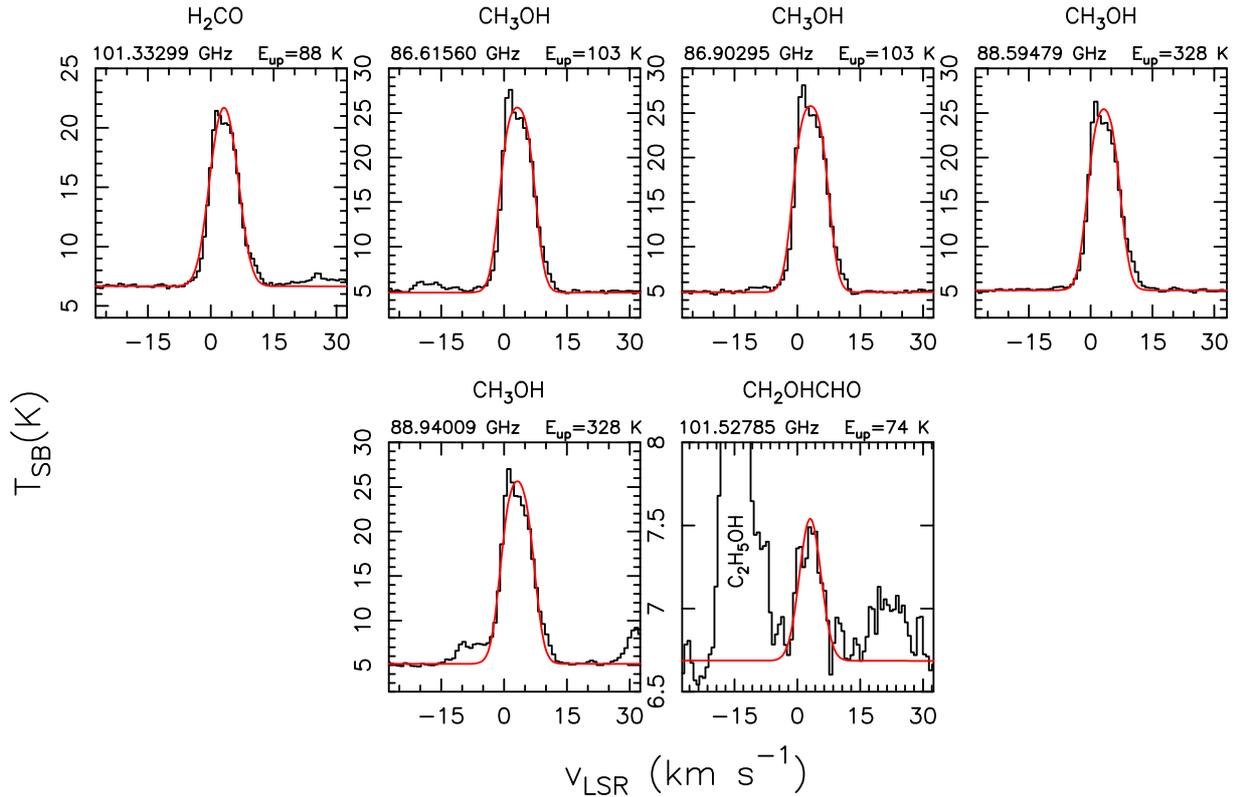}
 \caption{Unblended molecular transitions detected with ALMA of H$_2$CO, CH$_3$OH and CH$_2$OHCHO towards IRAS16293 source A. The MADCUBA-AUTOFIT LTE best fits, including the continuum, are shown in red. The frequency and $E_{\rm up}$ of each transition are indicated above each panel. Other molecular species in the spectra are also labeled with the their name or with an "U" if they remain unidentified.}
    \label{fig-molecules-A}
\end{figure*}

In Figures \ref{fig-molecules} and \ref{fig-molecules-A} we show the transitions of the other molecules studied in this work (H$_2$CO, CH$_3$OH and CH$_2$OHCHO; see Table \ref{table-transitions}) towards source B and A, respectively. The v$_{\rm LSR}$ and $FWHM$ of these species, shown in Table \ref{table-parameters}, is in good agreement with those of HCO in both sources. For source A, only the CH$_2$OHCHO transition at 101.53 GHz (lower panel in Figure \ref{fig-molecules-A}) is clearly unblended, while the others are contaminated by other species due to the larger linewidths.

We also did a careful search of the molecular lines that are close to those analysed in this paper in Figures \ref{fig-HCO}, \ref{fig-molecules} and \ref{fig-molecules-A}. To identify them, we searched for molecules using JPL and CDMS catalogs in a range of $\pm$1 MHz (which translates to $\pm$3.5 km s$^{-1}$) around the frequency of the lines. We successfully identified CH$_2$DOH, CH$_3$NCO (see Figure \ref{fig-HCO}), DCOOH, C$_2$H$_5$CN and HOCH$_2$CN  (Figure \ref{fig-molecules}), and C$_2$H$_5$OH (Figure \ref{fig-molecules}). Some lines in the spectra cannot be explained by any of the molecules from the catalogs, and then they remain unidentified (labeled with an ''U''). These lines might be due to species whose spectroscopy is still not available.




\begin{figure*}
\includegraphics[width=18cm]{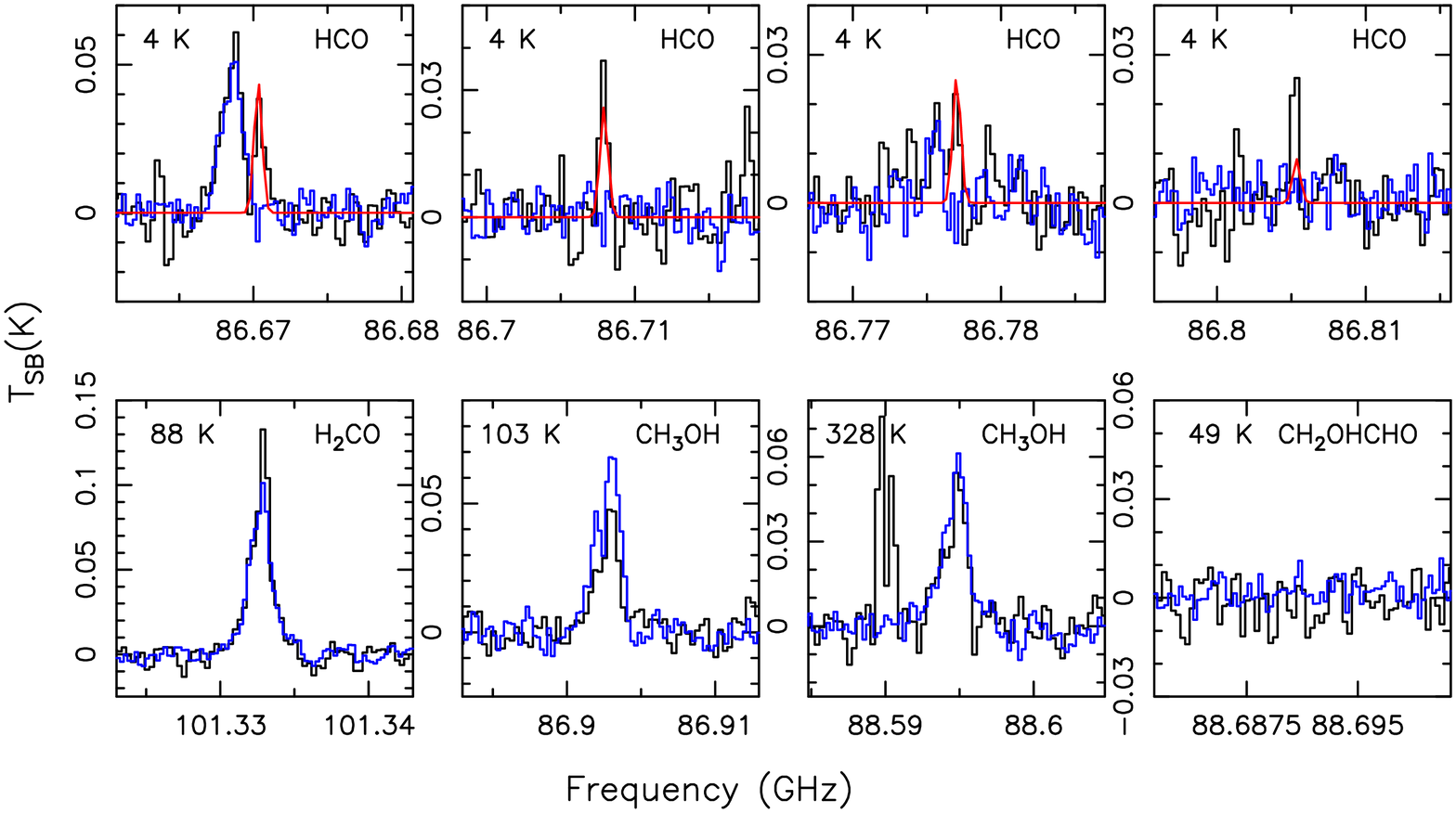}
 \caption{Comparison between IRAM 30m spectra (black histogram) and the ALMA spectra (blue histogram) integrated in a region coincident with the 30m beam. Different panels show different transitions. The molecule and the $E_{\rm up}$ are indicated in the right and left upper corners of each panel, respectively. The red curve in the upper panels show the LTE best fit for the HCO detected by the IRAM 30m telescope.}
    \label{fig-comparison}
\end{figure*}

\subsubsection{Comparison with single-dish data}  
\label{section-comparison}

We compared the ALMA spectra with the single-dish spectra observed with the IRAM 30m telescope, publicly available from the TIMASSS project (\citealt{caux2011}). This allows us to determine whether the ALMA observations have missed flux due to filtering, and also whether the single-dish were sensitive to gas located in the hot corinos or alternatively in a more extended and colder gas component surrounding the binary system.

In Figure \ref{fig-comparison}, we show the spectra of HCO, H$_2$CO, CH$_3$OH and CH$_2$OHCHO obtained with the IRAM-30m together with those of ALMA integrated in an area matching the IRAM 30m beam ($\sim$28$^{\prime\prime}$). As seen in this figure, the 3 mm quadruplet of HCO is also clearly detected in emission in the single-dish data (black histogram in upper panels).  On the other hand, HCO is not detected in the spatially averaged ALMA spectra (blue histogram in \ref{fig-comparison}; note that the line detected in the left upper panel is CH$_2$DOH). The emission/absorption detected by ALMA towards the hot corinos discussed previously is clearly diluted in the spatially-averaged spectra.  
This indicates that the HCO emission detected by the IRAM 30m telescope does not arise from the hot corinos but from a more extended component, probably associated with the circumbinary envelope (e.g. \citealt{van-dischoeck1995} and \citealt{quenard2018a}), which has been filtered-out by ALMA. We note that these ALMA observations had a maximum recoverable scale of $\sim$12$\arcsec$.


Unlike HCO, the lines of H$_2$CO and CH$_3$OH detected with the IRAM 30m telescope match very well the spatially averaged ALMA spectra, which indicates that the emission detected by the single-dish is arising from the compact hot corinos. 
As seen in Figure \ref{fig-comparison},  CH$_2$OHCHO, which has been clearly detected by ALMA (see Figures \ref{fig-molecules} and \ref{fig-molecules-A}),  is not detected when the emission is averaged over a 28$^{\prime\prime}$ area or when observed with the IRAM 30m due to beam dilution.




Altogether, we can conclude from this comparison that the extended HCO emission observed with the IRAM 30m comes from the bulk of the circumbinary envelope, and has been filtered out by ALMA, and that the molecular emission (HCO, H$_2$CO, and CH$_3$OH) detected with ALMA arises from the hot corinos. The HCO absorption profiles are due to foreground infalling gas of the common circumbinary envelope that is much colder than the hot corinos in the background.
This points out the need of using interferometric observations to study the HCO emission associated with the hot corinos/cores.

\begin{table*}
\tabcolsep 3pt
\begin{center}
\caption{Physical parameters obtained with the LTE analysis for the hot corinos IRAS16293 A \& B using the ALMA data, and for the cold envelope using IRAM 30m data. The parameters derived by using MADCUBA-AUTOFIT include their associated uncertainties, while the parameters that were fixed in the fitting process are shown without uncertainty. For the line opacities, we give the range of values found for the different molecular transitions of each species used in the analysis.}
\label{table-parameters}
\begin{tabular}{c c c c c c c c c c c}
\hline
Molecule    & $T_{\rm ex}$	   & v$_{\rm LSR}$	& FWHM & \multicolumn{2}{c}{$N$}     & &	 \multicolumn{2}{c}{Molecular abundance}    & $\tau$        & ${\overline{\chi^{2}}}_{\rm AUTOFIT}$\\ \cline{5-6}  \cline{8-9}
&    (K)       & (km s$^{-1}$) &    (km s$^{-1}$)  &  \multicolumn{2}{c}{($\times$ 10$^{16}$ cm$^{-2}$)}     &      &  \multicolumn{2}{c}{($\times$10$^{-8}$)}  & & \\ \cline{8-9} \cline{5-6}
& & & & this work & other works    	    	&   & this work & other works	   &     &   \\ 
\hline
\multicolumn{10}{c}{Source A (hot corino)} \\
\hline
HCO (emission) & 50  & 2.2 & 5.5 & 0.25 & - &  & 0.06 &- & 0.013$-$0.064  & - \\
                         & 180 & 2.2 & 5.5 & 1.6   & - &  & 0.36 &- & 0.003$-$0.017  &  - \\
HCO (absorption)  & 10  & 4.3 & 1 & 0.02$\pm$0.01$^{(a)}$ &  - &  & - & - & 0.2$-$0.97    & - \\
H$_2$CO & 100  & 3.2$\pm$0.1  & 7.1$\pm$0.2 &    15.2$\pm$0.4 & - & &  3.4$\pm$0.1  & 10$^{(b)}$  & 0.5  & 0.009 \\
CH$_3$OH   & 178$\pm$5  & 3.1$\pm$0.1 &  7.6$\pm$0.2 & 550$\pm$12 &  440 $^{(c)}$ & & 125$\pm$3 & 77$^{(c)}$ & 0.41$-$0.47   & 0.02 \\
CH$_2$OHCHO    & 110 & 3.1 & 7 & 1.1$\pm$0.2 & 4$^{(d)}$ &  & 0.26$\pm$0.03 & 0.75$^{(d)}$ & 0. 02  &  0.06 \\
                              & 140 & 3.1 & 7 & 1.5$\pm$0.2 & 4$^{(d)}$ &  & 0.34$\pm$0.04 & 0.75$^{(d)}$ & 0.016  & 0.06 \\
\hline
 \multicolumn{10}{c}{Source B (hot corino)} \\
\hline
HCO (emission)  & 50    & 2.5 & 1.8 & 0.5$\pm$0.2  & - & & 0.010$\pm$0.003 &- & 0.08$-$0.4  & 0.09\\
                           & 180  & 2.5 & 1.8 & 3.0$\pm$0.7  & - & & 0.06$\pm$0.01  &- & 0.017$-$0.083  & 0.09\\
HCO (absorption)  & 10  & 4.3$\pm$0.1  &  1 & 0.0046$\pm$0.0006  & - &  & - & -& 0.05$-$0.26  & 0.05\\
H$_2$CO & 100$^{(e)}$ & 2.4$\pm$0.1 & 1.3$\pm$0.1 & 23$\pm$3 & 200$^{(f)}$ &  & 0.46$\pm$0.06 & 8.6$^{(f)}$-10$^{(b)}$ & 4  & 0.006\\
CH$_3$OH     & 181$\pm$9 & 2.4$\pm$0.1 & 1.5$\pm$0.1 & 282$\pm$10 & 850$^{(b)}-$2000$^{(g)}$   & & 5.6$\pm$0.2 & 37$^{(c)}-$88$^{(g)}$ & 1.0$-$1.1  & 0.009 \\
CH$_2$OHCHO   & 157$\pm$14 & 2.7$\pm$0.1 & 1.8$\pm$0.2 & 13$\pm$3 & 3$-$6.8$^{(d,g)}$ &  & 0.27$\pm$0.05 & 0.16$-$0.35$^{(d,g)}$& 0.02$-$0.25  & 0.03 \\
\hline
 \multicolumn{10}{c}{Cold extended envelope} \\
\hline 
HCO & 10$^{(h)}$ & 4.0$\pm$0.3 & 2.2$\pm$0.5 & (2.6$\pm$0.6)$\times$10$^{-4}$  $^{(i)}$  & $^{(j)}$   &   & (0.13$\pm$0.03)$\times$10$^{-2}$  $^{(i)}$& -   & 0.001$-$0.006  & 0.002 \\

\hline
\end{tabular}
\end{center}
{(a) The AUTOFIT algorithm did not converge. We selected a good solution by visual inspection, and adopted a conservative uncertainty of 50$\%$.
(b)  \citet{ceccarelli2000}; 
(c) \citet{kuan2004}; 
(d)  \citet{jorgensen12}; 
(e) We have assumed a source size of 0.65$\arcsec$ for the the fit assuming T$_{\rm ex}$=100 K (see text); 
(f)  \citet{persson2017}; 
(g)  \citet{jorgensen2016}; 
(h) Temperature of the surrounding core of IRAS16293 of 10 K,  from \citet{van-dischoeck1995} and \citealt{quenard2018a};
(i) Column density averaged in the IRAM 30m beam;
(j) Using the hydrogen column density of the cold envelope reported by \citet{van-dischoeck1995}, $N$(H$_2$) = 2$\times$10$^{23}$ cm$^{-2}$.
}
%
\end{table*}

\subsubsection{Derivation of physical parameters}
\label{analysis-lines-parameters}

To estimate the physical parameters of the different species, we assumed Local Thermodynamic Equilibrium (LTE) conditions, which is a reasonably good approximation for hot corinos, where the volume densities are very high ($\geq$10$^8$ cm$^{-3}$; \citealt{jorgensen2016,coutens2018,quenard2018a}).  These densities are several orders of magnitude higher than the critical densities of the H$_2$CO and CH$_3$OH transitions considered in this work, which are $\sim$10$^4$ cm$^{-3}$, calculated from the collisional coefficients derived by \citet{wiesenfeld2013} and \citet{rabli2010}, respectively. For HCO and CH$_2$OHCHO, the LTE approach is the only possibility since there are no collisional coefficients available.

%

To fit the absorption profiles produced by the foreground molecular infalling layer, we used the expression:

\begin{equation*}
{
T_{\rm a}(\nu) =  f_{\rm c}(\nu) T_{\rm c}(\nu) e^{-\tau_{\rm a}(\nu)}   + f_{\rm a}(\nu) T_{\rm ex}^{\rm a}  (1-e^{-\tau_{\rm a}(\nu)})  , 
}
\end{equation*}
where $T_{\rm ex}$ is the excitation temperature of the absorbing molecular layer,  $\tau_{\rm e}(\nu)$ is the line optical depth of the absorption,  $f_{\rm c}(\nu)$ and $f_{\rm a}(\nu)$  are the beam filling factors of the continuum and the absorption regions, respectively, and  $T_{\rm c}(\nu)$ is the emission of the background continuum source described in Appendix \ref{continuum-fit}. This fit reproduces the absorption lines and also the continuum emission.

To fit the line intensities of the molecular emission from the hot corinos, $T_{\rm e}(\nu)$, we have used the expression: 

\begin{equation*}
{
T_{\rm e}(\nu) = f_{\rm e}(\nu) T_{\rm ex}^{\rm e} (1-e^{-\tau_{\rm e}(\nu)}), 
}
\end{equation*}
where $T_{\rm ex}$ is the excitation temperature of the emission, $\tau_{\rm e}(\nu)$ is the line optical depth of the emission, and $f_{\rm e}(\nu)$ is the beam filling factor of the emission region. Since the molecular emission and absorption and the continuum emission are not resolved towards both hot corinos, as discussed in Section \ref{analysis-lines-identification}, we considered the same beam dilution factor for all components, i.e., $f_{\rm c}(\nu)=f_{\rm e}(\nu)=f_{\rm a}(\nu)$, calculated using the continuum sizes presented previously: 0.88$\arcsec$ for source A and 0.44$\arcsec$ for source B.

The MADCUBA-AUTOFIT tool compares the observed spectra with the LTE synthetic spectra of the different species calculated following the previous expressions, taking into account all transitions, and it provides the best non-linear least-squared fit using the Levenberg-Marquardt algorithm.
The free parameters of each component (emission and absoprtion) are: total column density ($N$), excitation temperature ($T_{\rm ex}$), velocity ($\varv$), and full width half maximum ($FWHM$). MADCUBA-AUTOFIT calculates consistently from these parameters the line opacity for each transition between levels $i$ and $j$, $\tau_{\rm ij}$, using the expression: 

\begin{equation}
\tau_{\rm ij}= \frac{h}{FWHM} N_{\rm i} B_{\rm ij} (e^{h\nu_{\rm ij}/KT_{\rm ex}}-1),
\end{equation}
where {\it B}$_{\rm ij}$ is the Einstein $B$-coefficient, $h$ is the Planck constant, and $N_{\rm i}$ is the column density in the upper level $i$, which is calculated with the expression:

\begin{equation}
N_{\rm i}= \frac{N}{Q(T_{\rm ex})} g_{\rm i}  e^{- E_{\rm i}/KT_{\rm ex}},
\end{equation}
where $Q$ is the partition function and $g_{\rm i}$ and $E_{\rm i}$ are the degeneracy and the energy of the upper level $i$, respectively. The derived value of the opacity for each transition is given as output of the fit.

Letting free the 4 parameters ($N$, $T_{\rm ex}$, $\varv$, $FWHM$), MADCUBA-AUTOFIT  provides the best combination of parameters with the associated errors.
When the algorithm did not converge, we fixed manually the velocities and/or the FWHM to the values that best reproduced the observed spectra, and rerun AUTOFIT. In some cases,  the value of $T_{\rm ex}$ was also fixed (see below). When convergence was not possible, we selected by-eye the solution that best fits the spectra.
The physical parameters derived are shown in Table \ref{table-parameters}. The errors of the parameters left free are derived from the diagonal elements of covariance matrix, the inverse of the Hessian Matrix, and the final $\chi^2$ of the fit. 
The ratio between the errors and the values of each parameter gives a qualitative idea of the goodnes of the fit. Furthermore, in Table \ref{table-parameters}  we include the normalized $\overline{\chi^2}$ value of the fit, which was calculated dividing the $\chi^2$ by the signal-to-noise of the analysed transitions.




We were not able to derive the excitation temperature of HCO because the 4 detected lines of HCO share the same $E_{\rm up}$, 4 K. Since the ALMA maps (Figure \ref{fig-maps}) show that the HCO emission arises from the hot corinos (see also Section \ref{section-comparison}), we then assumed a range of temperatures typical of hot corinos, 50$-$180 K, which encompasses the values found for many other molecular tracers in the literature (e.g., \citealt{kuan2004,martin-domenech2017}) and the ones we obtain from CH$_3$OH and CH$_2$OHCHO in this dataset (see below).
The derived value of the HCO column density can vary up to a factor of 6 in the adopted temperature range. We will find in Section \ref{models} that the predictions of our chemical model favor a temperature for HCO of $\sim$70 K, within the range considered here.
For the HCO component in absorption, which is expected to arise from foreground and colder gas, we assumed the temperature of 10 K of the surrounding circumbinary envelope of IRAS16293  (\citealt{van-dischoeck1995} and \citealt{quenard2018a}).
The LTE fits of HCO in sources A and B, including both emission and absorption components, are shown with red curves in Figure \ref{fig-HCO}.
We note that while the brightest HCO transition at 86.670 GHz is detected in emission towards both hot corinos at detection levels >5$\sigma$, the other weaker transitions of the quadruplet are not clearly detected in emission. This might be due to a combination of several effects: weak line emission, noise of the spectra, modest spectral resolution, and the presence of the absorption profiles. To prove this, we simulated the HCO spectra (emission + absorption) using the physical parameters resulting from the fit (Table \ref{table-parameters}), mimicking the spectral resolution and the $rms$ of the observed spectra. We performed many different simulations in which the noise was created randomly. We also applied different spectral resamplings, i.e., we shifted slightly the position of the channels by values lower than the channel width.  Our results show that while the brightest HCO transition is always detected above 5$\sigma$, the other transitions can be spectrally diluted and fall below 2$-$3$\sigma$ levels, which would explain why they are not clearly detected in our observations. A similar spectral dilution effect of an inverse P-Cygni profile was observed in Herschel observations of the L1544 prestellar core with different spectral resolutions (\citealt{caselli&keto2012}).


The LTE fits of the other molecules are shown in Figures \ref{fig-molecules} (source B) and \ref{fig-molecules-A} (source A). 
For CH$_3$OH and CH$_2$OHCHO, for which multiple transitions with different $E_{\rm up}$ were observed (Table \ref{table-transitions}), we left $T_{\rm ex}^{e}$ as free parameter, and were able to derive its value. For CH$_3$OH, we obtained $T_{\rm ex}^{\rm e}$=178$\pm$5 K and 181$\pm$9 K for source A and B, respectively, which confirms that the methanol emission arises from the hot corinos.
For CH$_2$OHCHO, most of the unblended lines towards source B are low energy levels (Table \ref{table-transitions}). To avoid a bias towards low energies, we have selected 2 transitions at low energies and 2 transitions at high energies (indicated in Table \ref{table-transitions}) to perform AUTOFIT. We obtained a temperature of 157$\pm$14 K for source B, similar to that obtained for CH$_3$OH, and a column density of (13$\pm$3)$\times$10$^{16}$ cm$^{-2}$ (Table \ref{table-parameters}). 
For source A, the only clearly unblended transition of CH$_2$OHCHO is that at  101.527 GHz ($E_{\rm up}$= 74 K) . The excitation temperatures found in other studies for source A are typically a factor 0.7$-$0.9 lower than those of source B (\citealt{kuan2004,jorgensen12,jorgensen2016}). Therefore, we considered the temperature derived for source B and applied this range of factors, which gives $T_{\rm ex }^{\rm e}\sim$110$-$140 K. Fixing these temperatures, we fitted the CH$_2$OHCHO transition at 101.527 GHz, and we obtained column densities of (1.1$-$1.5)$\times$10$^{16}$ cm$^{-2}$ (Table \ref{table-parameters}).

The column density derived for CH$_3$OH in source B is 2.8$\times$10$^{18}$ cm$^{-2}$. We have  searched in the literature other estimates of the CH$_3$OH column density obtained from interferometric observations. For source B, \citet{kuan2004} reported a value of 5$\times$10$^{17}$ cm$^{-2}$ averaged in a synthesized beam of 1.2$\arcsec\times$2.6$\arcsec$. Considering the source size that we have adopted, 0.44$\arcsec$, this translates into $\sim$8.5$\times$10$^{18}$ cm$^{-2}$, which is about a factor of 3 higher than our value.  \citet{jorgensen2016} found a column density one order of magnitude higher than our value. The low value we found could be due to optical depth effects.  \citet{jorgensen2016} did not use CH$_3$OH but the isotopologue CH$_3^{18}$OH, which is expected to be optically thinner, and then used the standard ISM $^{16}$O/$^{18}$O ratio of 560 (\citealt{wilson&rood1994}). However,  the $^{16}$O/$^{18}$O ratio might be uncertain.
It has been found that other fractionation ratios, such as the $^{12}$C/$^{13}$C ratio, are usually lower in hot corinos/cores than in the ISM by a factor of $\sim$2 (\citealt{jorgensen2016}) or even of 4$-$5 (\citealt{beltran2018}). 
However, \citet{persson2017}
found a $^{16}$O/$^{18}$O ratio in source B that is not lower than the ISM value but higher, with a value of $\sim$805. Using this value, the CH$_3$OH column density derived by \citet{jorgensen2016} would be a factor 1.4 higher. Therefore, the most likely explanation for the low value we derived is optical depth. 
MADCUBA$-$SLIM, as explained before, calculates consistently the line opacity of each transition, and takes it into account in the calculation of the column density. Namely, if the source size and the $T_{\rm ex}$ are reasonably well known, the derived value of $N$ is good whenever the fitted lines are optically thin. If the lines are optically thick,
the derived $N$ should be considered as a lower limit, because the LTE line profile flattens and the line intensity becomes independent of $N$. 
The opacities derived with MADCUBA-AUTOFIT for the transitions of CH$_3$OH are 1.0$-$1.1 for source B (Table \ref{table-parameters}). Therefore, the lines are not optically thin, and our estimates of the CH$_3$OH column densities should be considered as strict lower limits.
Regarding source A, \citet{kuan2004} found a value of 4.4$\times$10$^{18}$ cm$^{-2}$ (after correction for the source size), which is very similar to our estimate of 5.5$\times$10$^{19}$ cm$^{-2}$. However, given the relatively high optical depth derived in our data (0.41$-$0.47, Table \ref{table-parameters}), these values should be considered as lower limits.




For H$_2$CO, we only have one transition, so $T_{\rm ex}^{\rm e}$ cannot be derived. We have used an excitation temperature of 100 K, like that found by  \citet{van-dischoeck1995} and \citet{persson2017}. We obtained a column density of (15.2$\pm$0.4)$\times$10$^{16}$ cm$^{-2}$ for source A. For source B, the LTE fit assuming $T_{\rm ex}^{\rm e}$=100 K saturates at a line intensity of $\sim$35 K, while the observed line temperature is $\sim$50 K (Figure \ref{fig-molecules}). This may indicate that the H$_2$CO emission is slightly larger than the 0.44$\arcsec$ size we have assumed. Considering a larger size of 0.65$\arcsec$ (still smaller than our beam), the line is well reproduced with a column density of (23$\pm$3)$\times$10$^{16}$ cm$^{-2}$, which is one order of magnitude lower than that found by \citet{persson2017} (Table \ref{table-parameters}). We note that the derived opacity of the line is high, $\sim$4, which indicates that this line is optically thick towards source B, and that therefore we are underestimating its column density.





To derive the molecular abundances of the different species we have used molecular hydrogen column densities calculated previously: N(H$_2$)=4.4$\times$10$^{24}$ cm$^{-2}$ and N(H$_2$)=5.0$\times$10$^{25}$ cm$^{-2}$ for A and B, respectively, 
The resulting molecular abundances are shown in Table \ref{table-parameters}. For source A, the HCO abundance is (0.06$-$0.36)$\times$10$^{-8}$,  for the range of excitation temperature considered. The molecular abundances obtained for the other species (H$_2$CO, CH$_3$OH and CH$_2$OCHO) are similar to previous estimates (Table \ref{table-parameters}) within factors 2$-$3. 

For source B, the HCO abundance is (0.01$-$0.06)$\times$10$^{-8}$, which is a factor of six lower than for source A. As discussed before,  CH$_3$OH may be suffering optical depth effects, which could explain why our abundance is a factor 5 and 16 lower than that estimated by \citet{kuan2004} and \citet{jorgensen2016}, respectively. The situation is similar for H$_2$CO, for which we have found an abundance more than one order of magnitude lower than that found by \citet{ceccarelli2000} and \citet{persson2017}. The abundance of CH$_2$OHCHO is very similar to that found by \citet{jorgensen12,jorgensen2016}.

\begin{table}
\tabcolsep 3pt
\begin{center}
\caption{Chemical models considered for the formation of HCO.} 
\label{table-formationHCO}
\begin{tabular}{c r}
\hline
\multicolumn{2}{c}{Model I}   \\
\hline
H$_2$CO + H$^+$  $\longrightarrow$ H$_3$CO$^+$ ; H$_3$CO$^+$ + e$^-$  $\longrightarrow$ HCO + H + H (gas) &   [1] \\
C + H$_2$O $ \longrightarrow$ HCO + H  (gas)   &  [2]\\
\hline

 \multicolumn{2}{c}{Model II}  \\
\hline
Model I \hspace{0.4cm}     +  \hspace{0.4cm}      CO $\longrightarrow$ HCO $\longrightarrow$ H$_2$CO $\longrightarrow$ CH$_3$OH (surface)    &  [3]\\
\hline

 \multicolumn{2}{c}{Model III}   \\
\hline
Model I     \hspace{0.4cm}       +      \hspace{0.4cm}    OH + H$_2$CO $\longrightarrow$ HCO + H$_2$O (surface)  &  [4] \\
\hline
\end{tabular}
\end{center}
\end{table}



We also derived the column density of the extended HCO component detected by the IRAM 30m telescope by fitting the HCO quadruplet shown in black in Figure \ref{fig-comparison}. We assumed that the emission fills the telescope beam, and used as $T_{\rm ex}$ the kinetic temperature of $\sim$10 K estimated for the  surrounding circumbinary envelope of IRAS16293 (\citealt{van-dischoeck1995} and \citealt{quenard2018a}).
The fit obtained with MADCUBA-AUTOFIT is shown in Figure \ref{fig-comparison}. We note that the observed intensities of the lines of the HCO quadruplet are not fully reproduced by the LTE fit. In particular, the lines at 86.70836 GHz and 86.80578 GHz are brighter than the LTE prediction, which may indicate that HCO in the cold circumbinary envelope is not in LTE. However, since collisional coefficients of this species are not available, we performed a LTE analysis. The derived parameters are shown in Table \ref{table-parameters}. We obtained a column density of (3$-$5)$\times$10$^{12}$ cm$^{-2}$, which translates into an abundance of (1.5$-$2.9)$\times$10$^{-11}$ by using the column density of the cold circumbinary envelope estimated by  \citet{van-dischoeck1995}, which is   $N_{\rm H_2}$=2$\times$10$^{23}$ cm$^{-2}$. This HCO abundance is significantly lower by 1$-$2 orders of magnitude than that found in the hot corinos.

\begin{figure*}
\includegraphics[angle=0,width=18cm]{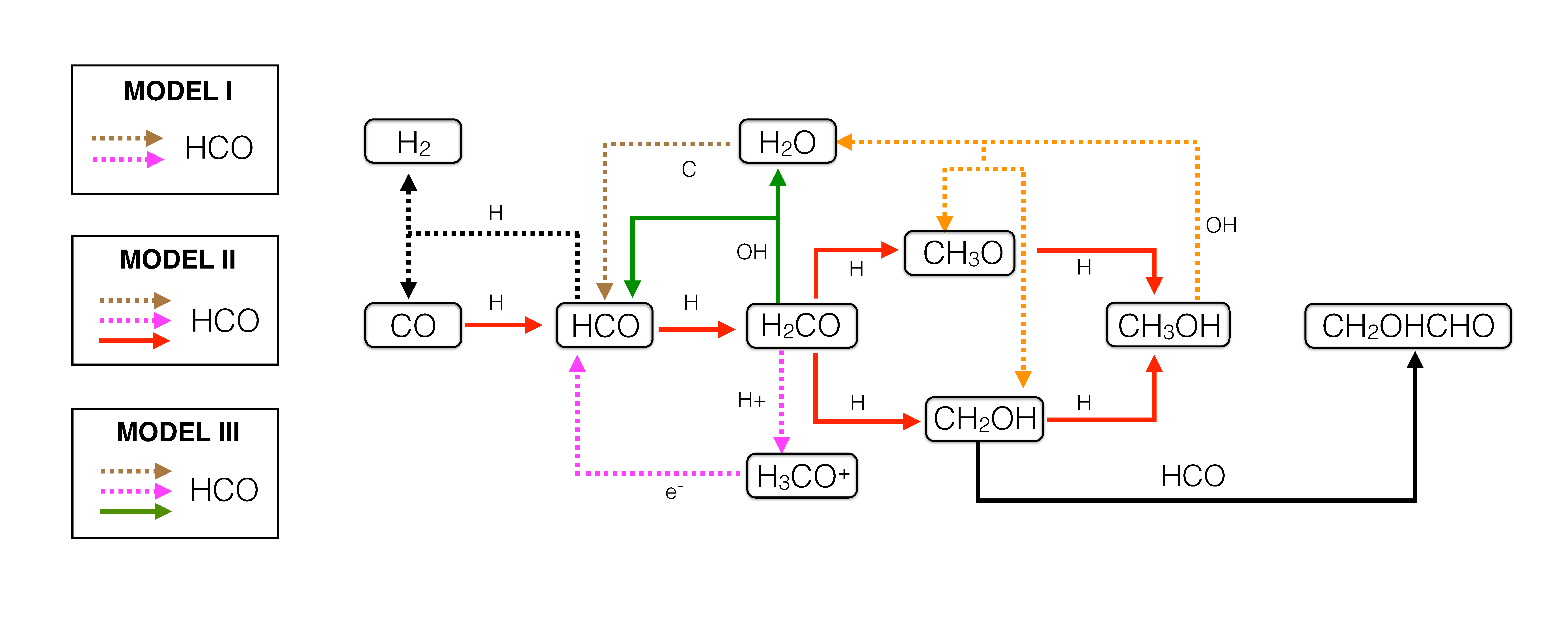}
 \caption{Chemical routes discussed in this work and implemented in our chemical model to study the formation and HCO. The solid and dashed arrows indicate gas-phase and grain surface reactions, respectively. On the left, we indicate with boxes the three different models (I, II and III) we considered for the formation of HCO. We indicate with arrows within each box the reactions that were included in each model, following the same color and stroke code than in the diagram. We have also added the most efficient formation route of CH$_2$OHCHO according to our model (Section \ref{GA-formation}).}
 \label{fig-HCO-diagram}
\end{figure*}

\section{Comparison with the chemical model}
\label{models}

To identify the most likely chemical routes to form HCO and its role in forming more complex species such as CH$_3$OH and CH$_2$OHCHO, we used an updated chemical model based on \citet{vasyunin2013b}. 
This schematic 0-dimensional model mimics the evolution of a parcel of gas and dust with time-dependent physical conditions from a diffuse cloud to a hot corino. The model consists of two stages. During the first stage, a free-fall collapse occurs (\citealt{brown1988,spitzer1998}). Physically, this stage is attributed to the formation of a cold and dense core from a translucent cloud. The collapse occurs over 10$^{6}$~yr. During this time, gas density increases from the initial value of 3$\times$10$^{3}$~cm$^{-3}$ to the final value of 10$^{8}$~cm$^{-3}$, which matches the typical densities of hot corinos (see e.g. \citealt{woods13,awad2014,coutens2018}). Visual extinction changes correspondingly from the starting value of $A_{\rm v}$=2, and gas and dust temperatures are assumed to be equal.
During the collapse, the dust temperature drops from $\sim$20~K to $\sim$10~K due to less efficient radiative heating of dust grains in a dark dense cloud in comparison to a translucent cloud exposed to UV radiation (\citealt{garrod&pauly2011}). 
The second stage is a warm-up phase. It is assumed that during this stage a parcel of gas and dust warms up from 10~K to 200~K during 2$\times$10$^{5}$~years developing into a hot corino. 
The gas density during the second stage remains constant. We would like to note that despite the simplicity of such model, it has been proven to be a powerful tool to explore chemistry of hot cores and corinos by a number of studies (see e.g. \citealt{brown1988,rawlings1992,viti1999,viti2004,garrod2006,vasyunin2013a,rivilla2016}). Thus we stick to this simplistic physical model, although more advanced treatment is planned for future detailed studies.

The chemical network used in this study is based on an updated version of the one presented in \citet{vasyunin2017}. Several important updates were introduced to accurately treat the chemistry of HCO and CH$_2$OHCHO. We note that the results of astrochemical models, including the one presented in this work, are intrinsically uncertain by at least an order of magnitude (\citealt{vasyunin2004,vasyunin2008, wakelam2005,wakelam2006, wakelam2010}). Thus, for the molecular abundances predicted by the model we have considered an uncertainty range multiplying and dividing by a factor of 3.

\subsection{The formation of HCO: surface chemistry, gas-phase chemistry or both?}
\label{HCO-formation}

Different chemical routes have been proposed to form HCO, both in the surface of dust grains (\citealt{tielens1982}; \citealt{brown1988}; \citealt{dartois1999}; \citealt{watanabe2002}; \citealt{woon2002}; \citealt{bacmann2016}) and in the gas-phase (\citealt{bacmann2016}; \citealt{hickson2016}).
The scheme of the chemical network including all these reactions  is presented in Figure \ref{fig-HCO-diagram}. To evaluate the role of surface chemistry in the formation of HCO, we have run the three different models (I, II and III) presented in Table \ref{table-formationHCO} (see also Figure \ref{fig-HCO-diagram}).
We also considered different reactions that can destroy HCO (see Figure \ref{fig-HCO-diagram}), both in gas phase and on grain surface.
As we will discuss in Section \ref{GA-formation}, HCO is involved in several chemical routes to form CH$_2$OHCHO.
For simplicity, in this section we only considered the route involving HCO and CH$_2$OH, which is the most efficient according to our model (see Section \ref{GA-formation}).





\begin{figure}
\includegraphics[width=8cm]{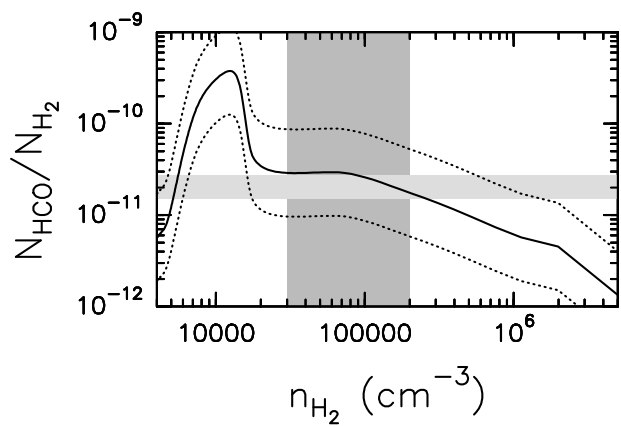}
 \caption{Results of the chemical model I (gas-phase chemistry) during the cold collapse phase. The evolution of the molecular abundance of HCO is represented as a function of the volume density with a solid black curve. The area within the dotted black curves delimit the uncertainty of the chemical model, obtained by multiplying and dividing the outcome of the model by a factor of 3. 
 The vertical dark gray band is the volume density derived by \citet{van-dischoeck1995} and \citet{quenard2018b} for the cold envelope of IRAS16293: 3$\times$10$^{4}-$ 2$\times$10$^{5}$ cm$^{-3}$. The  horizontal light gray band indicates the range of HCO abundances derived from the IRAM 30m observations (Table \ref{table-parameters}).}
\label{fig-hco-cold}
\end{figure}

The role of surface chemistry routes in determining the gas-phase abundance of HCO is different during the cold collapse phase and the warm-up phase. In the cold phase, the abundance of HCO is almost independent of the details of surface chemistry, and is mainly maintained by the gas-phase reaction [2] (Table \ref{table-formationHCO}).  Figure \ref{fig-hco-cold} shows that the abundance predicted by the chemical model I, which considers only gas-phase chemistry, is around several $\sim$10$^{-11}$ at the expected volume density of the cold envelope, 3$\times$10$^{4}-$2$\times$10$^5$ cm$^{-3}$ (from \citealt{van-dischoeck1995} and \citealt{quenard2018a}). This value is in good agreement with the abundance derived for the extended cold component  (see Table \ref{table-parameters}), indicating that  gas-phase formation routes are sufficient to consistently explain the HCO abundance in the cold envelope. This is also in agreement with \citet{bacmann2016} who reached the same conclusion in a survey of low-mass prestellar cores.





\begin{figure}
\includegraphics[width=8cm]{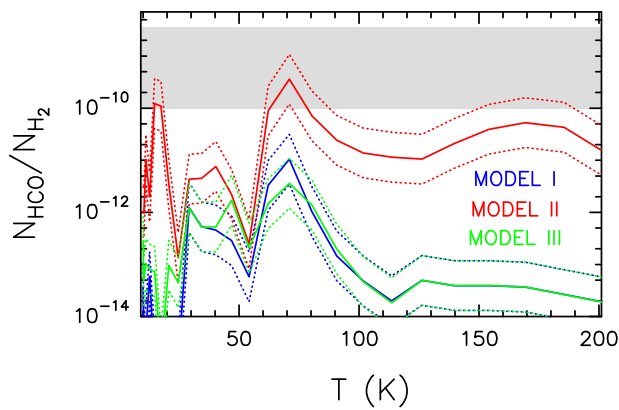}
 \caption{Results of the chemical models (I, II and III) during the protostellar warm-up phase. The evolution of the molecular abundance of HCO is represented as a function of temperature. The area within the dotted curves delimit the uncertainty of the chemical model, obtained by multiplying and dividing the output of the model  by a factor of 3. Model I (blue) includes only gas-phase reactions; model II (red) includes gas-phase reactions and the surface hydrogenation of CO; and model III (green) includes gas-phase reactions and the surface reaction H$_2$CO + OH $\rightarrow$ HCO + H$_2$CO. The horizontal light gray band indicates the range of HCO abundances found in the ALMA observations (Table \ref{table-parameters}).}
\label{fig-hco-model}
\end{figure}

During the protostellar phase, the contribution of grain-surface is much more important. In particular, the model II, which includes the surface hydrogenation of CO (reaction [3] in Table \ref{table-formationHCO}),  produces a peak of HCO abundance of several 10$^{-10}$, similar to that ones found in IRAS16293 A $\&$ B when a temperature of 50 K was assumed, which is higher than the abundances produced by Models I and III by more than an order of magnitude (Figure \ref{fig-hco-model}).
The peak of HCO abundance in the model is reached at a warm temperature of $\sim$60$-70$ K.
At higher temperatures, HCO is progressively destroyed through the gas-phase reaction between HCO and atomic hydrogen (\citealt{hebrard2009}). This reaction is very efficient in hotter gas because the abundance of atomic hydrogen is significantly higher than in cold gas. 
Therefore, the chemical model is suggesting that the emission of HCO detected by ALMA is arising for a warm component of the hot corinos at a temperature $\sim$60$-$70 K, rather than from hotter gas ($>$100 K). Since we were not able to derive the excitation temperature of HCO, in Section \ref{results} we assumed two temperatures, 50 and 180 K, to derive its abundance. Indeed, the HCO abundances derived assuming 50 K, which are a factor $\sim$6 lower, match better the peak abundances of HCO predicted by the model (at 60$-$70 K). This is a further evidence pointing towards a warm origin of the HCO observed in emission by ALMA.

\begin{figure*}
\includegraphics[width=18cm]{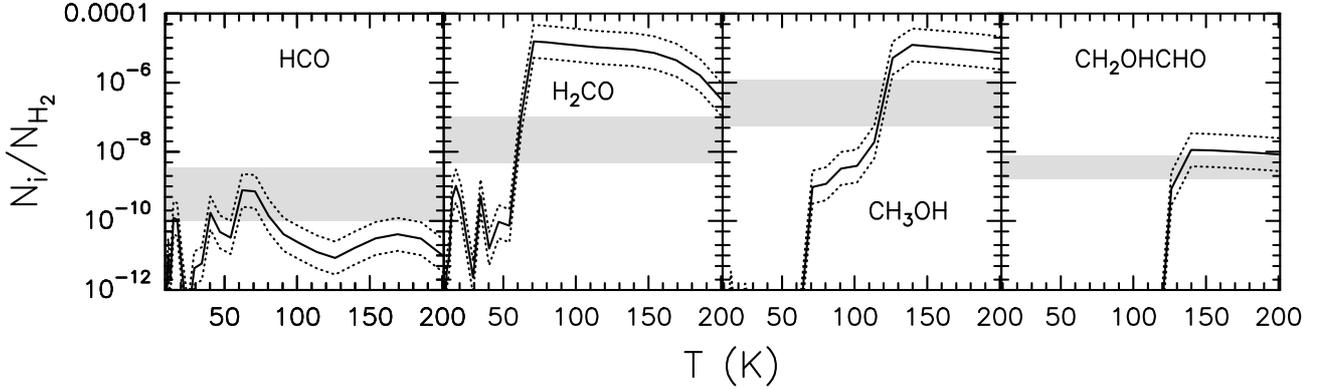}
 \caption{Results of the chemical model that better fits the molecular abundances of the species studied in this work. We show the molecular abundances as a function of the temperature with solid curves. The area within the dotted curves delimit the uncertainty of the chemical model, obtained multiplying and dividing the model result by a factor of 3. The horizontal light gray bands indicate the range of abundances found in the observations for source A and B (Table \ref{table-parameters}).
}
    \label{fig-models}
\end{figure*}

\subsection{HCO as precursor of COMs}
\label{GA-formation}

HCO has been proposed as the basic precursor of many COMs, and in particular of CH$_2$OHCHO.
\citet{woods12,woods13} tested theoretically different mechanisms of CH$_2$OHCHO synthesis previously proposed in the literature (e.g. \citealt{sorrell01,Charnley_Rodgers2005,halfen2006,bennett07,garrod08,beltran09}), both in the gas phase and on the surface of grains, and concluded that the most likely pathways are three grain-surface formation routes involving HCO. 
To explore the possible formation routes of CH$_2$OHCHO, we considered several routes recently discussed in the literature (\citealt{garrod08, fedoseev15,chuang2016, woods12,woods13}. In all these routes, the availability of HCO is a key condition. 
All routes of formation of HCO considered in the previous section were switched on.

Therefore, to test the viability of these chemical pathways to form CH$_2$OHCHO, we have compared the molecular abundances estimated in IRAS16293 with the output of the considered models.
 Since the detailed modeling of CH$_{2}$OHCHO chemistry will be presented in a separate paper (Vasyunin et al., in prep.), below we only focus on the chemical pathway that showed the best agreement with observations presented in this study.

The model that better fits the observations is the model where CH$_2$OHCHO is produced by the surface reaction between HCO and CH$_2$OH. It produces a peak of HCO of $\sim$ 0.07$\times$10$^{-8}$ at 60$-$70 K, which is of the order of that derived for sources A $\&$ B if we assume $T_{\rm ex}$=50 K (see left panel of Figure \ref{fig-models}).  For CH$_3$OH, the observed abundances are reasonably well reproduced for temperatures $\geq$110 K  (Figure \ref{fig-models}), which are not far from the excitation temperature derived from observations (see Table \ref{table-parameters}). 

The abundance of H$_2$CO derived from observations is reached at a temperature of $\sim$60 K in the model (Figure \ref{fig-models}), similar to the temperature at which the abundance of HCO peaks according to the model. This may indicate that HCO and H$_2$CO are tracing warm gas of the hot corinos, instead of the hotter gas ($T>$100 K) traced by CH$_3$OH and CH$_2$OHCHO. Alternatively, if H$_2$CO traces hot gas at $\sim$100 K, as suggested by the excitation temperature found by \citet{van-dischoeck1995} and \citet{persson2017}, then the model is overproducing H$_2$CO by $\sim$2 orders of magnitude compared to observed values. 

Finally, CH$_{2}$OHCHO in the model reaches an abundance similar to that estimated from the observations for a temperature of $\sim$140~K, which is consistent with the $T_{\rm ex}$ derived from the observations. This temperature is higher than that at which HCO peaks in the model ($\sim$70~K). 
This is because CH$_{2}$OHCHO, which is formed at low temperatures through the surface reaction HCO+CH$_{2}$OH$\rightarrow$CH$_{2}$OHCHO, has much higher desorption energy than HCO (6684~K vs. 1600~K; \citealt{garrod08}), and then it is desorbed later when the temperature increases.

In summary, we favor the reaction between HCO and CH$_2$OH as the most likely chemical route for the formation of CH$_2$OHCHO. However,  we do not aim at completely rule out other surface-chemistry routes recently considered in laboratory experiments (\citealt{fedoseev15,chuang2016}) and chemical modeling (\citealt{woods12,woods13,coutens2018}). A more complete and detailed analysis of all routes of formation of CH$_{2}$OHCHO, including also the recently proposed gas-phase route by \citet{skouteris2018} will be presented in a forthcoming paper (Vasyunin et al., in prep.).

\section{Summary and conclusions}

The formyl radical HCO has been proposed as the fundamental precursor of many complex organic molecules. We presented the first high spatial resolution maps of HCO using ALMA towards the Solar-type protostellar binary IRAS16293$-$2422. We also detected several lines of the chemically related species formaldehyde (H$_2$CO), methanol (CH$_3$OH) and glycolaldehyde (CH$_2$OHCHO). 
The HCO emission is compact and arises from the hot corinos surrounding the two protostars. The line profiles exhibit also redshifted absorption produced by foreground infalling material of the circumbinary cold envelope. Previous IRAM 30m single-dish observations detected more extended HCO emission from the circumbinary envelope. To identify the most likely chemical routes to form HCO, and to determine its role in forming more complex species, we compared the observed molecular abundances with the predictions of our chemical model, which takes both gas-phase and grain-surface chemistry into account. We find that while HCO in the envelope can be expained considering only pure gas-phase reactions, the HCO detected in the hot corinos needs the hydrogenation of CO on the surface of dust grains to be formed. Subsequent surface hydrogenation of HCO and thermal desorption are also responsible for the observed abundances of H$_2$CO and CH$_3$OH. We also find that the main formation route of CH$_2$OHCHO is the surface reaction between HCO and CH$_2$OH. The detailed contribution of other chemical routes, including gas-phase reactions, will be carefully studied in a forthcoming theoretical paper.


\section*{Acknowledgments}
We thank the anonymous referee for her/his constructive comments that contributed to improve the manuscript.
This project has received funding from the European Union's Horizon 2020 research and innovation programme under the Marie Sk\l{}odowska-Curie grant agreement No 664931, and from the Italian  Ministero  dell'Istruzione, Universit\`a e Ricerca through the grant Progetti Premiali 2012 - iALMA (CUP C52I13000140001).
This paper makes use of the following  ALMA data:
\noindent
ADS/JAO.ALMA\#2015.1.01193.S 
ALMA is a partnership of ESO (representing its  member states), NSF (USA) and  NINS (Japan), together with NRC (Canada), NSC and ASIAA (Taiwan), and KASI (Republic of Korea), in co-operation with the Republic of Chile. The Joint ALMA Observatory is operated by ESO, AUI/NRAO and NAOJ.

\bibliographystyle{mnras}
\bibliography{Bib} 

\clearpage
\appendix

\section{Gaussian fits of the molecular transitions}
\label{appendix-line-parameters}

We present in Tables \ref{table-transitions-gaussianA} and \ref{table-transitions-gaussianB} the results of the Gaussian fits of the different molecular species observed with ALMA towards source A and B, respectively. Table \ref{table-transitions-timasss} presents the results for the HCO quadruplet observed with the IRAM 30m telescope.

\begin{table*}
\tabcolsep 5.0pt
\begin{center}
\caption{Line parameters measured in the ALMA spectra obtained from Gaussian fits of the different molecular transitions towards IRAS16293-2422 source A.}
\label{table-transitions-gaussianA}
\begin{tabular}{l c c c c c c}
\hline
Molecule &  Frequency       & Transition	                &  Area 	               & Intensity    		& v$_{\rm LSR}$    & FWHM         \\ 
                & (GHz)                &                                   & (K km s$^{-1}$) & (K)                     &  (km s$^{-1}$)     &  (km s$^{-1}$)  \\ 
\hline
HCO (emission)     &  86.67076     &  1$_{0,1}-$0$_{0,0}$, J=3/2--1/2, F=2--1  &  5.8$\pm$0.8 & 1.1$\pm$0.1  & 1.9$\pm$0.2 & 4.9$\pm$0.6\\
HCO (absorption)     &  86.70836     &  1$_{0,1}-$0$_{0,0}$, J=3/2--1/2, F=1--0  & -1.3$\pm$0.6 & -1.0$\pm$0.5 & 4$\pm$1  & 0.9$\pm$0.7 \\
\hline
H$_2$CO & 101.33299     &  6$_{1,5}-$6$_{1,6}$     &  108$\pm$4    & 14.2$\pm$0.4  & 3.14$\pm$0.08  &  7.2$\pm$0.2 \\
\hline
CH$_3$OH & 86.61560     &  7$_{2,6}-$6$_{3,3}$ - -     & 182$\pm$7  & 22.2$\pm$0.5  & 3.24$\pm$0.08  &  7.7$\pm$0.2 \\
CH$_3$OH & 86.90295     &  7$_{2,5}$- 6$_{3,4}$ + +     &   185$\pm$6   & 22.7$\pm$0.5 & 3.24$\pm$0.08  &   7.7$\pm$0.2 \\
CH$_3$OH & 88.59479     &  15$_{3,13}$ - 14$_{4,10}$ ++    &    154$\pm$9  &  19.8$\pm$0.7 &  3.1$\pm$0.2 & 7.3$\pm$0.3  \\
CH$_3$OH & 88.94009     &  15$_{3,12}$ - 14$_{4,11}$ - -     &   166$\pm$8  &  20.5$\pm$0.6 & 3.3$\pm$0.2  &  7.6$\pm$0.3 \\
\hline
CH$_2$OHCHO & 101.52785   &    14$_{5,9}$ - 14$_{4,10}$    &  3.9$\pm$0.6   & 0.68$\pm$0.07   &   2.9$\pm$0.3 & 5.5$\pm$0.6 \\
\hline
\end{tabular}
\end{center}
\end{table*}

\begin{table*}
\tabcolsep 5.0pt
\begin{center}
\caption{Line parameters measured in the ALMA spectra obtained from Gaussian fits of the different molecular transitions towards IRAS16293-2422 source B.}
\label{table-transitions-gaussianB}
\begin{tabular}{l c c c c c c}
\hline
Molecule &  Frequency       & Transition	                &  Area 	               & Intensity    		& v$_{\rm LSR}$    & FWHM         \\ 
                & (GHz)                &                                   & (K km s$^{-1}$) & (K)                     &  (km s$^{-1}$)     &  (km s$^{-1}$)  \\ 
\hline
HCO (emission)     &  86.67076     &  1$_{0,1}-$0$_{0,0}$, J=3/2--1/2, F=2--1  & 5$\pm$3  & 2.3$\pm$0.8  & 2.4$\pm$0.5  & 2.3$\pm$0.7 \\
HCO (absoprtion)     &  86.67076     &  1$_{0,1}-$0$_{0,0}$, J=3/2--1/2, F=2--1  & -5$\pm$2 &  -4$\pm$1 &  4$\pm$1 & 1.0$\pm$0.7\\
HCO  (absoprtion)     &  86.70836     &  1$_{0,1}-$0$_{0,0}$, J=3/2--1/2, F=1--0  & -3.2$\pm$0.8  & -2.2$\pm$0.4 &  4.4$\pm$0.1 & 1.4$\pm$0.3 \\
HCO   (absoprtion)   &  86.77746     &  1$_{0,1}-$0$_{0,0}$, J=1/2--1/2, F=1--1     &   -3.3$\pm$0.5   & -2.8$\pm$0.3  &  4.2$\pm$0.1 & 1.1$\pm$0.2 \\
\hline
H$_2$CO & 101.33299     &  6$_{1,5}-$6$_{1,6}$     &  54$\pm$3   &  28.4$\pm$0.7 & 2.43$\pm$0.03  & 1.79 $\pm$0.05 \\
\hline
CH$_3$OH & 86.61560     &  7$_{2,6}-$6$_{3,3}$ - -     &  23$\pm$2 &  13.1$\pm$0.7 & 2.48$\pm$0.05  &  1.64$\pm$0.09 \\
CH$_3$OH & 86.90295     &  7$_{2,5}$- 6$_{3,4}$ + +     &   24$\pm$3     &   13.4$\pm$0.8 &  2.49$\pm$0.05  & 1.7$\pm$0.1    \\
CH$_3$OH & 88.59479     &  15$_{3,13}$ - 14$_{4,10}$ ++    &  22$\pm$2    & 12.6$\pm$0.5  &  2.38$\pm$0.03 & 1.64$\pm$0.08 \\
CH$_3$OH & 88.94009     &  15$_{3,12}$ - 14$_{4,11}$ - -     &   23$\pm$2  &  12.9$\pm$0.5  &   2.76$\pm$0.03 & 1.66$\pm$0.07 \\
\hline
CH$_2$OHCHO & 86.60057      &   17$_{5,2}$ - 17$_{4,13}$    &   6.2$\pm$0.9  & 3.9$\pm$0.4   & 2.60$\pm$0.07   &  1.5$\pm$0.2 \\
CH$_2$OHCHO &  86.86239     &  7$_{4,3}$ - 7$_{3,4}$    &   6$\pm$1  &  3.1$\pm$0.4   &  2.7$\pm$0.1    & 1.7$\pm$0.3 \\
CH$_2$OHCHO &  86.87650     &    $20_{4,16}$ - 20$_{3,17}$      & 2.9$\pm$0.7  & 1.9$\pm$0.3  & 2.58$\pm$0.08  & 1.4$\pm$0.3   \\ 
CH$_2$OHCHO & 88.53041      &   8$_{4,5}$ - 8$_{3,6}$    &   2.9$\pm$0.7   & 1.9$\pm$0.3   &  2.58$\pm$0.08   & 1.4$\pm$0.3 \\
CH$_2$OHCHO & 88.69126    &   12$_{3,10}$ - 12$_{2,11}$     &   6$\pm$2    & 3.7$\pm$0.6 & 2.8$\pm$0.1   & 1.4$\pm$0.3 \\
CH$_2$OHCHO & 88.89245    &   9$_{4,6}$ - 9$_{3,7}$     &   5$\pm$1     &  3.4$\pm$0.5  & 2.82$\pm$0.08  &  1.4$\pm$0.2  \\
CH$_2$OHCHO &  99.06847   &   14$_{4,11}$ - 14$_{3,12}$     &  6$\pm$2  & 5.1$\pm$0.7  & 2.61$\pm$0.07  & 1.2$\pm$0.2 \\ 
CH$_2$OHCHO & 101.11631   &   21$_{4,17}$ - 21$_{3,18}$     &  7$\pm$2  &  4.3$\pm$0.7 & 2.5$\pm$0.2  & 1.6$\pm$0.3  \\
CH$_2$OHCHO &  101.21981     &   $18_{3,15}$ - $18_{2,16}$     & 1.5$\pm$0.4     & 0.9$\pm$0.2 & 2.67$\pm$0.09  & 1.6$\pm$0.2 \\
CH$_2$OHCHO & 101.23217   &   15$_{2,13}$ - 15$_{1,14}$     &   6$\pm$2   & 3.8$\pm$0.6 &  2.6$\pm$0.2 &  1.5$\pm$0.3 \\
CH$_2$OHCHO &  101.51469     &   $14_{3,12}$ - $14_{2,13}$     & 1.3$\pm$0.5  & 0.6$\pm$0.2 &  2.3$\pm$0.2 & 1.9$\pm$0.5 \\
CH$_2$OHCHO & 101.52785   &    14$_{5,9}$ - 14$_{4,10}$    & 7.6$\pm$0.9   &  4.1$\pm$0.3  &  2.53$\pm$0.06 &  1.7$\pm$0.2 \\
\hline
\end{tabular}
\end{center}
\end{table*}

\begin{table*}
\tabcolsep 5.0pt
\begin{center}
\caption{HCO line parameters measured in the IRAM 30m spectra obtained from Gaussian fits.}
\label{table-transitions-timasss}
\begin{tabular}{l c c c c c c}
\hline
Molecule &  Frequency       & Transition	                &  Area 	               & Intensity    		& v$_{\rm LSR}$    & FWHM         \\ 
                & (GHz)                &                                   & (K km s$^{-1}$) & (K)                     &  (km s$^{-1}$)     &  (km s$^{-1}$)  \\ 
\hline
HCO     &  86.67076     &  1$_{0,1}-$0$_{0,0}$, J=3/2--1/2, F=2--1  &  0.07$\pm$0.04  &  0.03$\pm$0.01 & 3.6$\pm$0.4  & 2.0$\pm$0.7 \\
HCO     &  86.70836     &  1$_{0,1}-$0$_{0,0}$, J=3/2--1/2, F=1--0  & 0.06$\pm$0.03 & 0.03$\pm$0.01 &  4.2$\pm$0.4 & 1.8$\pm$0.6 \\
HCO     &  86.77746     &  1$_{0,1}-$0$_{0,0}$, J=1/2--1/2, F=1--1     &  0.03$\pm$0.02    &  0.016$\pm$0.008 & 4.3$\pm$0.5  &  1.6$\pm$0.8 \\
HCO     &  86.80578     &  1$_{0,1}-$0$_{0,0}$, J=1/2--1/2, F=0--1     &   0.03$\pm$0.02   &  0.02$\pm$0.01 & 4.4$\pm$0.5  & 1.1$\pm$0.9  \\
\hline
\end{tabular}
\end{center}
\end{table*}

\clearpage

\section{Fit of the continuum level}
\label{continuum-fit}

To reproduce the continuum level of the spectra we have used a modified black body function:

\begin{equation*}
{ 
T_{\rm c}(\nu) = f_{\rm c}(\nu)  B_{\rm \nu}(T_{\rm c})(1 - e^{-\tau(\nu)})= }
\end{equation*}
\begin{equation}
{
f_{\rm c}(\nu)  \frac{h\nu}{k} \frac{1}{e^{h\nu/kT_{\rm c}}-1} (1 - e^{-\tau(\nu)}) , 
}
\label{eq-gray-body}
\end{equation}
%
%
where {\it B}($T_{\rm c}$) is the blackbody function with temperature $T_{\rm c}$, $f_{\rm c}(\nu)$ is the beam filling factor of the continuum emission in each spectral window, and $\tau_{\rm d}(\nu)$=$\tau_{\rm d0}(\nu/\nu_{0})^{\beta}$ is the dust optical depth, where $\beta$ is the dust emissivity spectral index.
We adopted $\nu_{0}$=94 GHz (the intermediate frequency of the four spectral windows), and assumed that $T_{\rm c}$ is equal to the excitation temperature derived from CH$_3$OH, 178 K and 181 K for sources A and B respectively (see Section \ref{analysis-lines-parameters}).  
To calculate the beam filling factor we used the synthesized beams of the datacubes of the different spectral windows (Table \ref{table-observations}) and the deconvolved sizes of the continuum sources of A and B at 94 GHz (Figure \ref{fig-maps}): 1.1$\arcsec\times$0.7$\arcsec$ and 0.46$\arcsec\times$0.42$\arcsec$, respectively. 

\begin{figure*}
\includegraphics[width=16cm]{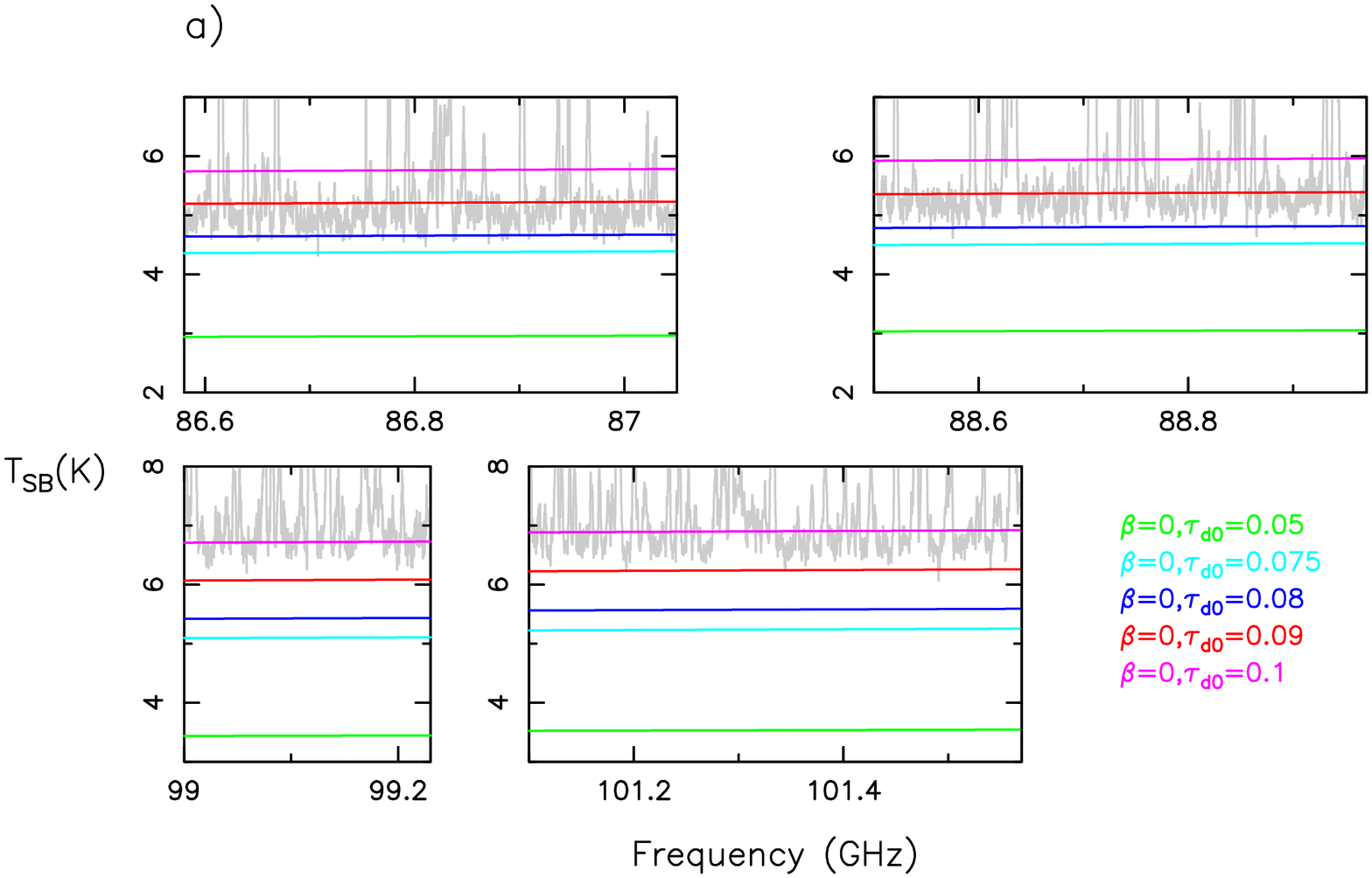}
\includegraphics[width=16cm]{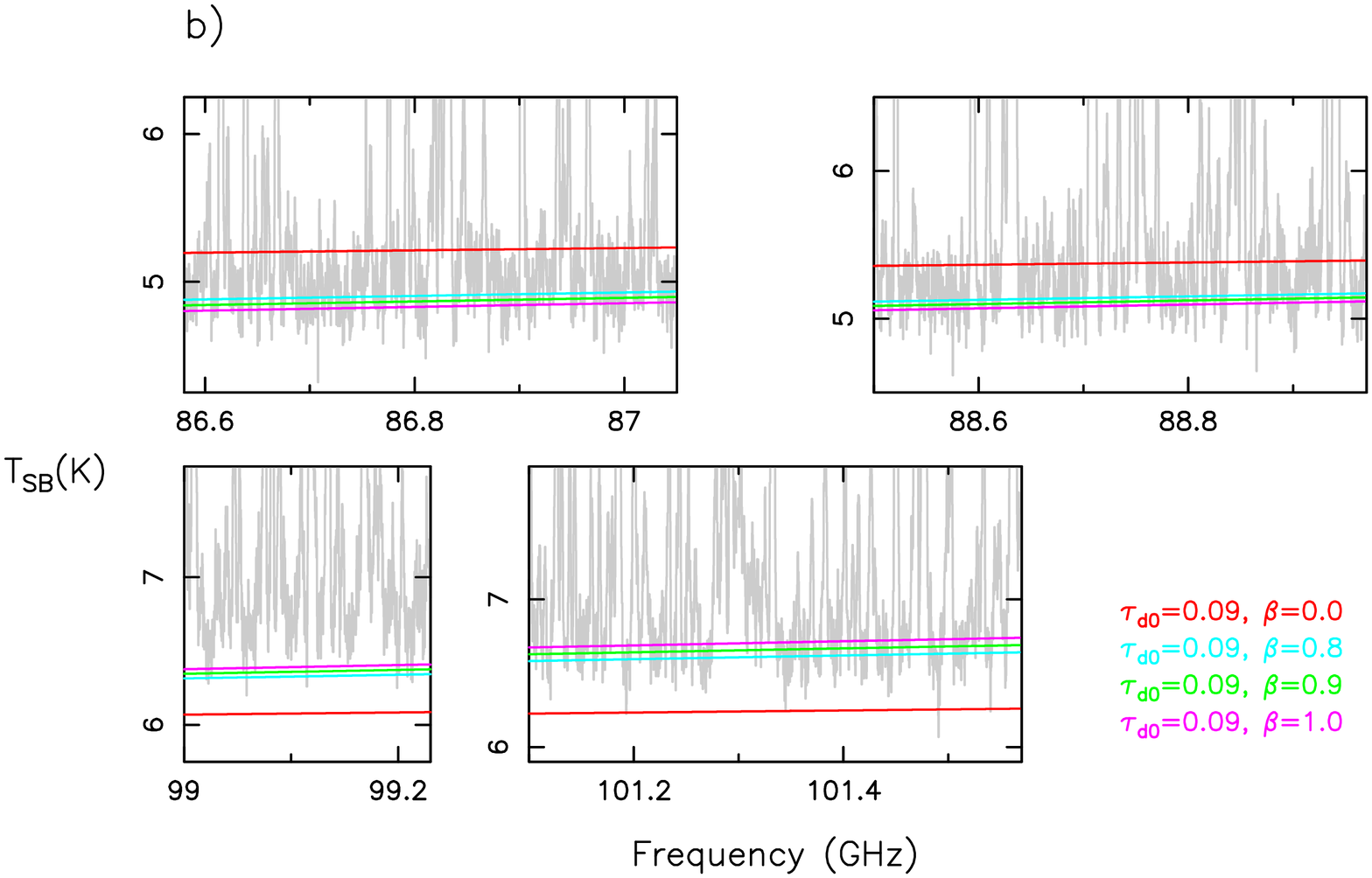}
\caption{Fit of the continuum level in the spectra towards the continuum peak of IRAS16293 A. The four different spectral windows are shown. The colored lines show different fits with different  pairs of values  of $\tau_{\rm d0}$ and $\beta$. We adopted $\nu_{\rm 0}$=94 GHz and $T_{\rm c}$=178 K. To calculate $f_{\rm c}$ we used a size of 1.1$\arcsec\times$0.7$\arcsec$. {\it Panel a):} we show different fits keeping $\beta$=0 and varying the value of $\tau_{\rm d0}$. {\it Panel b):} We show different fits keeping $\tau_{\rm d0}$=0.09 and varying the value of $\tau_{\rm d0}$.}
 \label{fig-continuoA}
\end{figure*}

\begin{figure*}
\includegraphics[width=16cm]{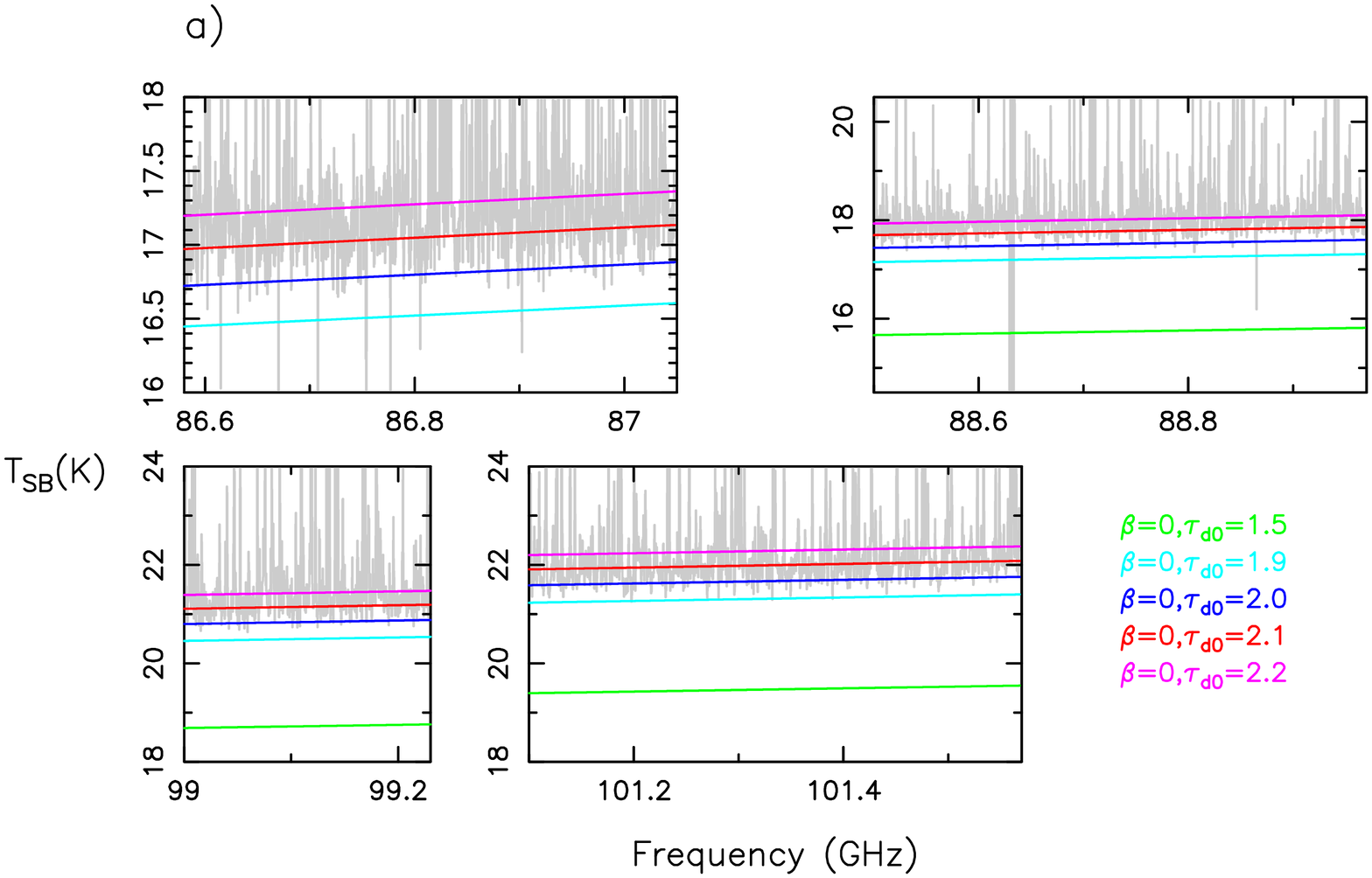}
\includegraphics[width=16cm]{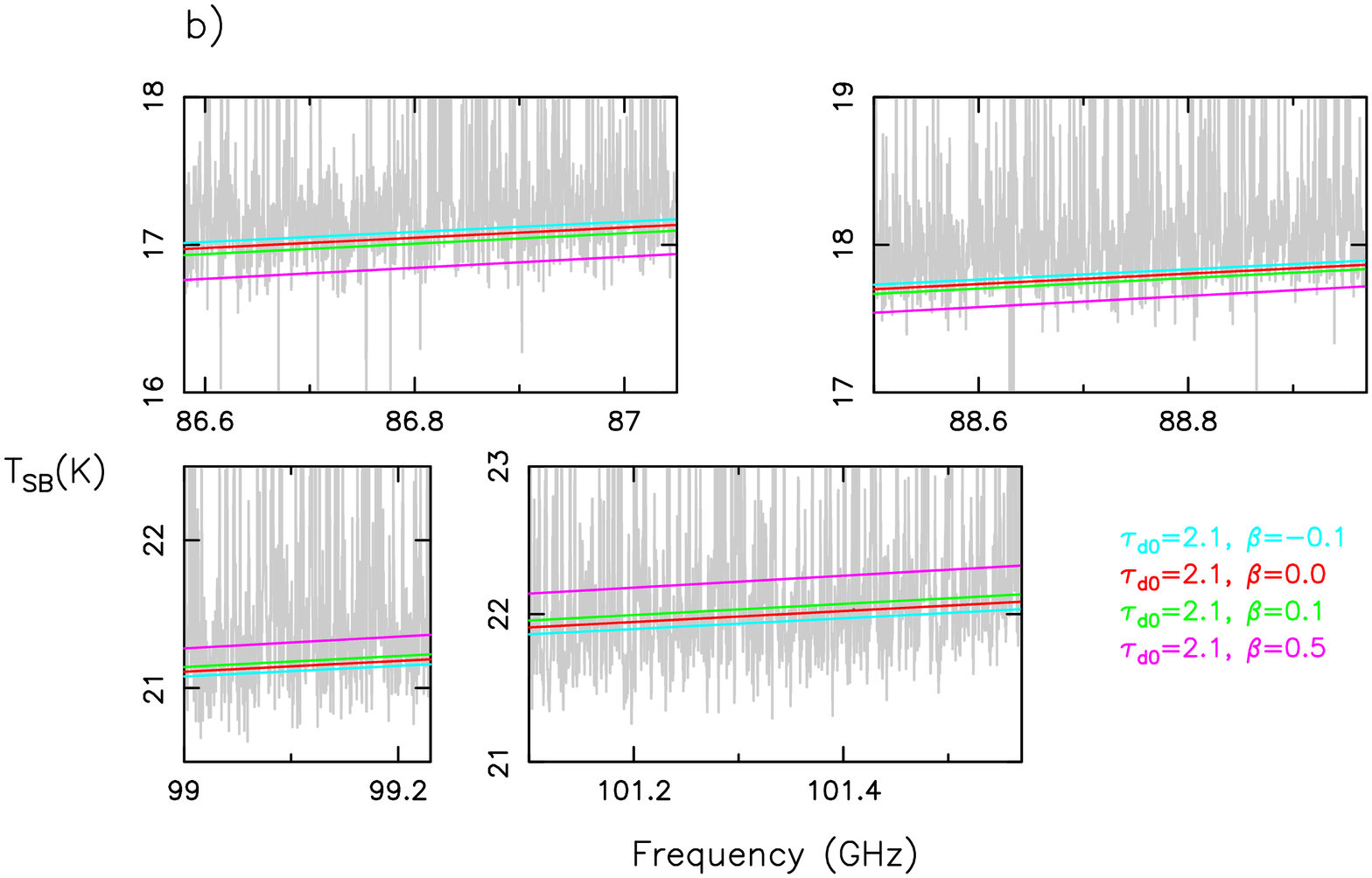}
\caption{Fit of the continuum level in the spectra towards the continuum peak of IRAS16293 B. The four different spectral windows are shown. The colored lines show different fits with different pairs of values  of $\tau_{\rm d0}$ and $\beta$. We adopted $\nu_{\rm 0}$=94 GHz and $T_{\rm c}$=181 K. To calculate $f_{\rm c}$ we used a size of 0.46$\arcsec\times$0.42$\arcsec$. { \it Panel a):} We show different fits keeping $\beta$=0 and varying the value of $\tau_{\rm d0}$.  {\it Panel b):} We show different fits keeping $\tau_{\rm d0}$=2.1 and varying the value of $\tau_{\rm d0}$.}
 \label{fig-continuoB}
\end{figure*}

We fitted the continuum emission of both hot corinos in all spectral windows by varying the values of  $\beta$ and $\tau_{d0}$.  The dust emissivity spectral index $\beta$ is the responsible of the behavior as a function of the frequency, i.e., the slope of the continuum emission, while the dust optical depth $\tau_{d0}$ controls the level of the continuum emission. We tried multiple combinations of both parameters (some examples are shown in Figures \ref{fig-continuoA} and \ref{fig-continuoB}) and chose by visual inspection the solutions that fit better the continuum level of the four spectral windows in each hot corino. Although the determination of the continuum level is not univocal due to the presence of multiple lines in such a crowded spectra, we note that the uncertainty from the visual inspection is lower than the rms of the spectra and that the uncertainty of the flux calibration, which we used to derive the errors of $\tau_{d0}$ and $\beta$. The fits that better reproduce the continuum level in all spectral windows are: $\beta$=0.9 and $\tau_{d0}$=0.09 for source A, and $\beta$=0 and  $\tau_{d0}$=2.1 for source B (see Figures \ref{fig-continuoA} and \ref{fig-continuoB}). We used these parameters to fit the continuum level in the Section \ref{analysis-continuum}.


We assumed a flux density calibration error of $\pm$5$\%$, from \citet{jorgensen2016} and \citet{bonato2018}. According to this, we expect calibration uncertainties of around $\pm$0.3 and $\pm$1 K for source A and B, respectively. We added to these values the rms of the spectra (0.15 K and 0.18 K, respectively), obtaining a total uncertainty in the continuum fluxes of  $\pm$0.45 K and  $\pm$1.18 K, respectively. Considering this, the values of  $\tau_{d0}$  with their associated uncertainties are 0.09$\substack{+0.01 \\ -0.01}$ and 2.1$\substack{+0.6 \\ -0.4}$ for source A and B, respectively.  
To estimate the error of $\beta$ we only considered the rms of the spectra and not the calibration uncertainty, since $\beta$ is giving the behavior with frequency and the calibration of the four spectral windows was the same. We then obtained for $\beta$: 0.9$\pm$0.4 and 0.0$\pm$0.5 for source A and B, respectively.

\bsp	
\label{lastpage}
\end{document}